\documentclass[aps,twocolumn,pra,eqsecnum, floats, showpacs, notitlepage, superscriptaddress,10pt]{revtex4-1}
\usepackage{graphicx}
\usepackage{dcolumn}
\usepackage{amsmath}
\usepackage{verbatim}
\usepackage{bbold}
\usepackage{subeqnarray}
\usepackage{bm}
\usepackage{color}
\usepackage{lmodern}
\usepackage[caption=false]{subfig}
\usepackage{amssymb}
\usepackage{mathrsfs}
\usepackage{pict2e}
\usepackage{tikz}
\usepackage{mathtools}

\DeclarePairedDelimiterX\braket[2]{\langle}{\rangle}{#1 \delimsize\vert #2}
\usepackage[breaklinks=true,colorlinks=true,linkcolor=blue,urlcolor=blue,citecolor=magenta]{hyperref}

 \newcounter{multifig}
 
\begin{document}
\title{Effect of spontaneous emission on a tanh model}

\author{A. D. Kammogne}
\affiliation{Mesoscopic and Multilayers Structures Laboratory, Faculty of Science, Department of Physics, University of Dschang, Cameroon}
\date{\today}

\begin{abstract}
This study examines the impact of spontaneous emission on a tanh model. The effect is characterized by introducing an imaginary term and a shift in the model, resulting in its non-Hermitian nature. This leads to the appearance of light beams in the population evolution, primarily influenced by the coupling strength and the shift. We derive the necessary conditions to identify the allowed and forbidden regions in the diagram of the real part of the energy. Additionally, we analyze how sweep velocity and time affect the imaginary part of the energy. Furthermore, we demonstrate the similarities between our model and the Rabi and Landau-Zener models. Throughout this work, we confirm that our theoretical predictions align well with numerical simulations.
\end{abstract}
\maketitle

\section*{Introduction}\label{Sec1}

The interaction between certain quantum systems and a laser or magnetic field can lead to the creation of spontaneous emission. This phenomenon underlies many physical processes, including dissipation between quantum systems \cite{Halbertal, Feynman, MHarrington, Celeghini, YYan, FernandoGalve, Tarantelli, Barontini, Schuch, Maddox}, the emergence of non-Hermitian properties \cite{ShaolinKe, AESiegman, AdiPick, XZZhang, Zongping, Marani, Pichler, Zhong, Siwei, Motohiko}, and the decay processes \cite{Anastopoulos, Xue-Hua, Poenaru, EJHellund, Grimsmo, AOBarut} during these interactions. Such phenomena are typically observed at atomic and subatomic scales. To further explore spontaneous emission, we utilize the tanh model, commonly referred to as the Demkov-Kunike model. This model is well-suited for studying this phenomenon due to its realistic representation of parameter evolution with a hyperbolic tangential behavior and its description of excitations involving fluorescence transitions from excited states to other states outside the system. The tanh model has been the focus of extensive scientific investigation.

For instance, the tanh model has been applied to study spontaneous decay in the work of Kenmoe et al.\cite{Kenmoe}, where Demkov-Kunike models were analyzed in the context of irreversible decay of ground and excited states. Spontaneous decay is incorporated by introducing an imaginary term in the diagonal part of the Hamiltonian, which accounts for the non-Hermitian nature of the system.

The tanh model is also used to explore the noise and the authors \cite{LinChen} shows the impact of telegraph and gaussian noise in the Demkov-Kunike model, they remark that the survival probability is suppressed for the slow telegraph noise while for a slow Gaussian noise, the noise always enhance the survival probability.

The second Demkov-Kunike model analyzed in \cite{Ishkhanyan}  is studied in a framework where the asymptotic solutions are expressed in terms of the corresponding quasi-energy. It is shown that resonance occurs asymmetrically. The tanh model is highly useful for describing the molecule formation in a Bose-Einstein condensate under the weak interaction regime, as demonstrated in the work of R. S. Sohhoyan et al. \cite{Sokhoyan}.

\begin{figure}[h!]
	\centering
	\includegraphics[width=0.52\textwidth, height =0.52\textwidth]{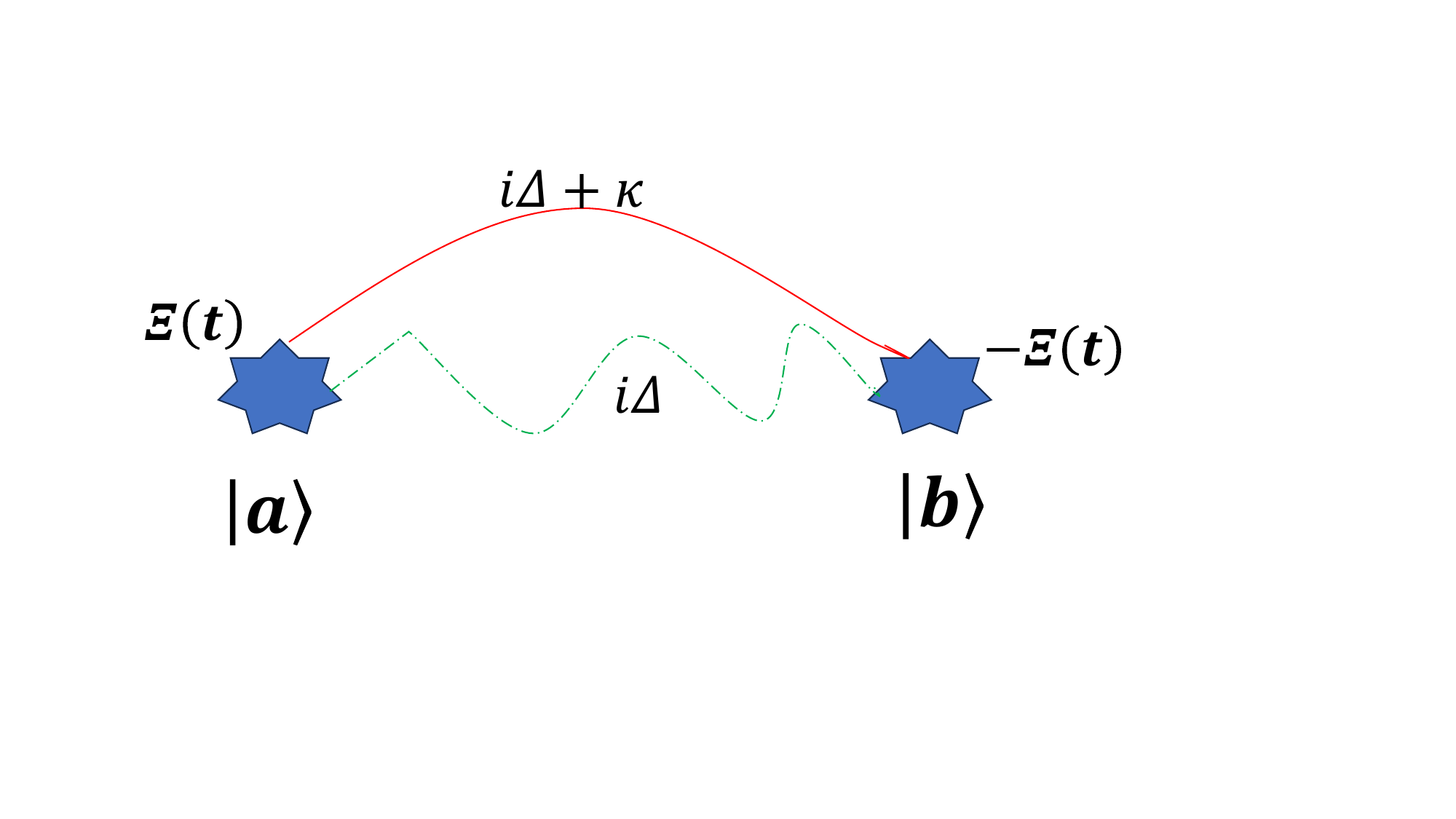}
	\caption{\small Illustration of the two-state Tanh model with spontaneous emission. In this figure, the states $\left| a \right\rangle$ and $\left| b \right\rangle$ are associated with the detuning $\Xi \left( t \right)$ and $-\Xi \left( t \right)$. The injection of a magnetic field induces spontaneous emission, which is characterized by an imaginary coupling term and a subsequent energy shift. This introduces a non-Hermitian character to our Hamiltonian.}. \label{fig1}
\end{figure}

Ping Feng et al.\cite{FengP}, analyze the nonlinear system in the context of the tanh model. They show that the transition probability can rapidly stabilize for the initial state $\psi_1$.  In the case of strong interaction, the quantum transition is completely suppressed. N. V. Vitanov and S. Stenhlom \cite{NVitanov} studied the tanh model under the introduction of a pulse excitation transitioning to a decaying level. They demonstrated that population loss decreases for higher decay rates.

In this paper, we propose a novel adaptation of the tanh model, incorporating the effects of spontaneous emission via the inclusion of an imaginary coupling term $-i\Delta$ and a shift $\kappa$ as illustrated in figure \ref{fig1}. Each of the state $a$ and $b$ correspond to the detuning $\Xi\left(t\right)$ and $-\Xi\left(t\right)$.  This spontaneous emission arises from the injection of a magnetic field, resulting in the non-Hermitian nature of our system. This property is particularly significant for metastable states, as spontaneous emission plays a crucial role in radioactive processes \cite{DBIon, Sandulescu, OAPTavares, Caitlin, MGOLDHABER, MMirea} and ultra cold trapped gas \cite{HerwigOtt, JMHutson, Bloch, JianpingYin, MTBell}.

In this work, the author construct a simplified theoretical framework to understand the dynamics of this model, ensuring the validity of both analytical and numerical solutions. The primary objective is to determine the transition probabilities of the system. Additionally, the secondary objectives include evaluating energy diagrams.

To clarify the methodology, we first present our model in Section \ref{sec2}. In Section \ref{sec3}, we describe the mathematical methods. Section \ref{sec4} focuses on the results and their discussion. Section \ref{sec5} is dedicated to the description of eigenenergies. In Section \ref{sec6}, we establish analogies with well-known models, such as the Rabi and Landau-Zener models, and conclude with a brief summary and potential perspectives.

\section{Tanh Model}\label{sec2}

The model studied to describe the effect of spontaneous emission here is the "tanh model," similar to the first Demkov-Kunike model but renormalized. Its Hamiltonian is given by:
\begin{equation}
	H\left( t \right) = \frac{1}{2} \left(\Xi \left( t \right){\sigma _z} + \Theta  {\sigma _x}\right) ,\label{1.1}
\end{equation}

where $\Xi \left (t \right)$ and $\upsilon $ represent the detuning and the Rabi frequency, respectively. A special feature of this model is the hyperbolic tangential behavior of the detuning. Spontaneous emission introduces an imaginary coupling term and a shift between states in the off-diagonal part of the system. For our model, these parameters are given by:
\begin{equation}
	\Xi \left( t \right) = P\tanh \left( {\alpha t + \beta } \right) + \kappa, \label{1.2}
\end{equation}

\begin{equation}
	\Theta = i\Delta  + \kappa. \label{1.3}
\end{equation}

We choose the sweep rate $\alpha$ such that  is positive, as the phase $\beta$ plays a major role in the oscillations observed in the system. Here, $P$ represents the population amplitude, while $\Delta$ is the imaginary coupling due to spontaneous emission in the system. The shift $\kappa$   is considered as a real coupling in the off-diagonal part and as a static component in the diagonal part of the Hamiltonian. The pauli matrices $\sigma _z$ and $\sigma _x$ help to polarize our system in z and x-directions. Our contribution to this model is the addition of a phase in the hyperbolic tangential part of our detuning and an imaginary part with a jump. This model is suitable for describing metastable states, decay processes in quantum mechanics, and quantum information processing. Having briefly presented our model, we'll now study the system dynamics under the effect of spontaneous emission.

\section{Mathematical method}\label{sec3}

The system dynamics are described using the Schr\"odinger picture through a time-dependent equation \cite{Kenmoe, Kammogne1, Kammogne2}. Since our system is non-Hermitian due to the presence of an imaginary coupling, this equation is given by:
\begin{equation}
	i\frac{d}{{dt}}B\left( t \right) = H\left( t \right)B\left( t \right),
\end{equation}

where the amplitude probabilities are represented by $B\left( t \right) = {\left[ {{\psi _1}\left( t \right),{\psi _2}\left( t \right)} \right]^T}$ and the Hamiltonian $H\left( t \right)$ is given in \eqref{1.1}. Using a gauge transformation ${B_1}\left( t \right) = {\psi _1}\left( t \right)\exp \left( { - \frac{i}{2}\int\limits_{{t_0}}^t {\Theta \left( {{t_1}} \right)d{t_1}} } \right)$, the general form of the differential equation is:
\begin{equation}
	\frac{{{d^2}{\psi _1}\left( t \right)}}{{d{t^2}}} - i\Xi \left( { t } \right)\frac{{d{\psi _1}\left( t \right)}}{{dt}} + \frac{1}{4}{{\Theta } ^2}{\psi _1}\left( x \right) = 0,\label{2.2}
\end{equation}

with aid of the change of variable $x\left( t \right) = \frac{1}{2}\left( {1 + \tanh \left( {\alpha t + \beta } \right)} \right)$, the differential equation transforms into:
\begin{equation}
	x\left( {1 - x} \right)\frac{{{d^2}{\psi _1}\left( x \right)}}{{d{x^2}}} + \left( {a - bx} \right)\frac{{d{\psi _1}\left( x \right)}}{{dx}} + \frac{{{c^2}}}{{x\left( {1 - x} \right)}}{\psi _1}\left( x \right) = 0,\label{2.3}
\end{equation}

where the parameters $a$, $b$, and $c$ are defined as:
\begin{equation}
	a = 1 + \frac{i}{{2\alpha }}\left( {P - \kappa } \right),\label{2.4}
\end{equation}
\begin{equation}
	b = 2\left( {1 + \frac{{iP}}{{2\alpha }}} \right),\label{2.5}
\end{equation}
\begin{equation}
	c = \frac{{i\Delta  + \kappa }}{{4\alpha }}.\label{2.6}
\end{equation}

This new system configuration highlights the role of parameter $a$  in amplitude and state jumps, parameter $b$  in amplitude adjustments, and parameter $c$ in observing spontaneous emission effects through the imaginary coupling.

Considering the ansatz transformation ${\psi _1}\left( x \right) = {x^\mu }{\left( {1 - x} \right)^\nu }z\left( x \right)$ for the wavefunction $\psi_1\left( x \right)$, the differential equation takes the form of the hypergeometric equation:

\begin{equation}
	x\left( {1 - x} \right)\frac{{{d^2}z\left( x \right)}}{{d{x^2}}} + \left( {\gamma  - \left( {\alpha  + \omega  + 1} \right)x} \right)\frac{{dz\left( x \right)}}{{dx}} + \rho \omega z\left( x \right) = 0,\label{2.7}
\end{equation}

where the parameters $\gamma, b $, and $\mu$ are given by:
\begin{equation}
	{\mu _{1,2}} = \frac{{1 - a \mp \sqrt {{{\left( {1 - a} \right)}^2} - 4{c^2}} }}{2},\label{2.8}
\end{equation}
\begin{equation}
	{\nu _{1,2}} = \frac{1}{2}\left( {1 + a - b \mp \sqrt {{{\left( {1 + a - b} \right)}^2} - 4{c^2}} } \right),\label{2.8a}
\end{equation}
\begin{equation}
	\rho  = \mu  + \nu,\label{2.8b}
\end{equation}
\begin{equation}
	\omega  = \mu  + \nu  + b - 1,\label{2.8c}
\end{equation}
\begin{equation}
	\gamma  = 2\mu  + a. \label{2.9}
\end{equation}

To preserve the symmetry of our problem, the equations \eqref{2.7} - \eqref{2.9} are similar to those of Kenmoe et al. \cite{Kenmoe}.Having presented the method required to study the dynamics of our system, we can move on to the next step, which consists of presenting our various results.

\section{Results}\label{sec4}

In this section, we derive expressions for the amplitude probabilities, the propagator, and the transition probability between states $a$ and $b$.

The equation \eqref{2.7} is solved using two independent solutions, expressed in terms of hypergeometric functions $M\left( {\mu ,\gamma ,z} \right)$ and $U\left( {\mu ,\gamma ,z} \right)$. By applying the ansatz transformation , the probabilities amplitude takes the form:
\begin{equation}
	{B_1}\left( {x,{x_0}} \right) = {x^\chi }{\left( {1 - x} \right)^\varsigma }\left( {{a_ + }\left( {{x_0}} \right){R_1}\left( x \right) + {a_ - }\left( {{x_0}} \right){T_1}\left( x \right)} \right), \label{2.10}
\end{equation}

\begin{equation}
	{B_2}\left( {x,{x_0}} \right) = {x^\chi }{\left( {1 - x} \right)^\varsigma }\left( {{a_ + }\left( {{x_0}} \right){R_2}\left( x \right) + {a_ - }\left( {{x_0}} \right){T_2}\left( x \right)} \right). \label{2.11}
\end{equation}

With

\begin{equation}
	{a_ \pm }\left( {{x_0}} \right) = {a_ \pm }x_0^{ - \chi }{\left( {1 - {x_0}} \right)^{ - \varsigma }},
\end{equation}
\begin{equation}
	{R_1}\left( x \right) = {x^\mu }{\left( {1 - x} \right)^\nu }F\left( {\rho ,\omega ,\gamma ,x} \right),
\end{equation}
\begin{equation}
	{T_1}\left( x \right) = {x^{1 + \mu  - \gamma }}{\left( {1 - x} \right)^\nu }F\left( {\rho  - \gamma  + 1,\omega  - \gamma  + 1,2 - \gamma ,x} \right),
\end{equation}
\begin{equation}
	\begin{gathered}
		{R_2}\left( x \right) = \frac{{4i\rho }}{\Delta }{x^{\mu  + 1}}{\left( {1 - x} \right)^{\nu  + 1}}(\frac{{\rho \beta }}{\gamma }F\left( {\rho  + 1,\omega  + 1,\gamma  + 1,x} \right) \hfill \\
		\,\,\,\,\,\,\,\,\,\,\,\,\,\,\,\,\,\,\,\,\,\,\,\,\, + \left( {\frac{{\mu  - \left( {\mu  + \nu } \right)x}}{{x\left( {1 - x} \right)}}} \right)F\left( {\rho ,\omega ,\gamma ,x} \right)) \hfill \\ 
	\end{gathered},
\end{equation}
\begin{equation}
	\begin{gathered}
		{T_2}\left( x \right) = \frac{{4i\alpha }}{\Delta }{x^{\mu  - \gamma  + 1}}{\left( {1 - x} \right)^{\nu  + 1}} \times  \hfill \\
		(\left( {\frac{{\mu  - \left( {\mu  + \nu } \right)x}}{{x\left( {1 - x} \right)}}} \right)F\left( {\rho  - \gamma  + 1,\omega  - \gamma  + 1,2 - \gamma ,x} \right) \hfill \\
		\,\,\,\,\,\,\,\,\,\,\,\,\,\,\,\,\,\,\,\,\,\,\,\,\, + \left( {1 - \gamma } \right)F\left( {\rho  - \gamma  + 1,\omega  - \gamma  + 1,1 - \gamma ,x} \right)) \hfill \\ 
	\end{gathered},
\end{equation}
\begin{equation}
	\chi  = \frac{{iA - i\kappa }}{{4\alpha }},
\end{equation}

where
\begin{equation}
	\varsigma  = \frac{{iA + i\kappa }}{{4\alpha }},
\end{equation}

\begin{figure}[h!]
	\centering
	\includegraphics[width=0.23\textwidth, height =0.25\textwidth]{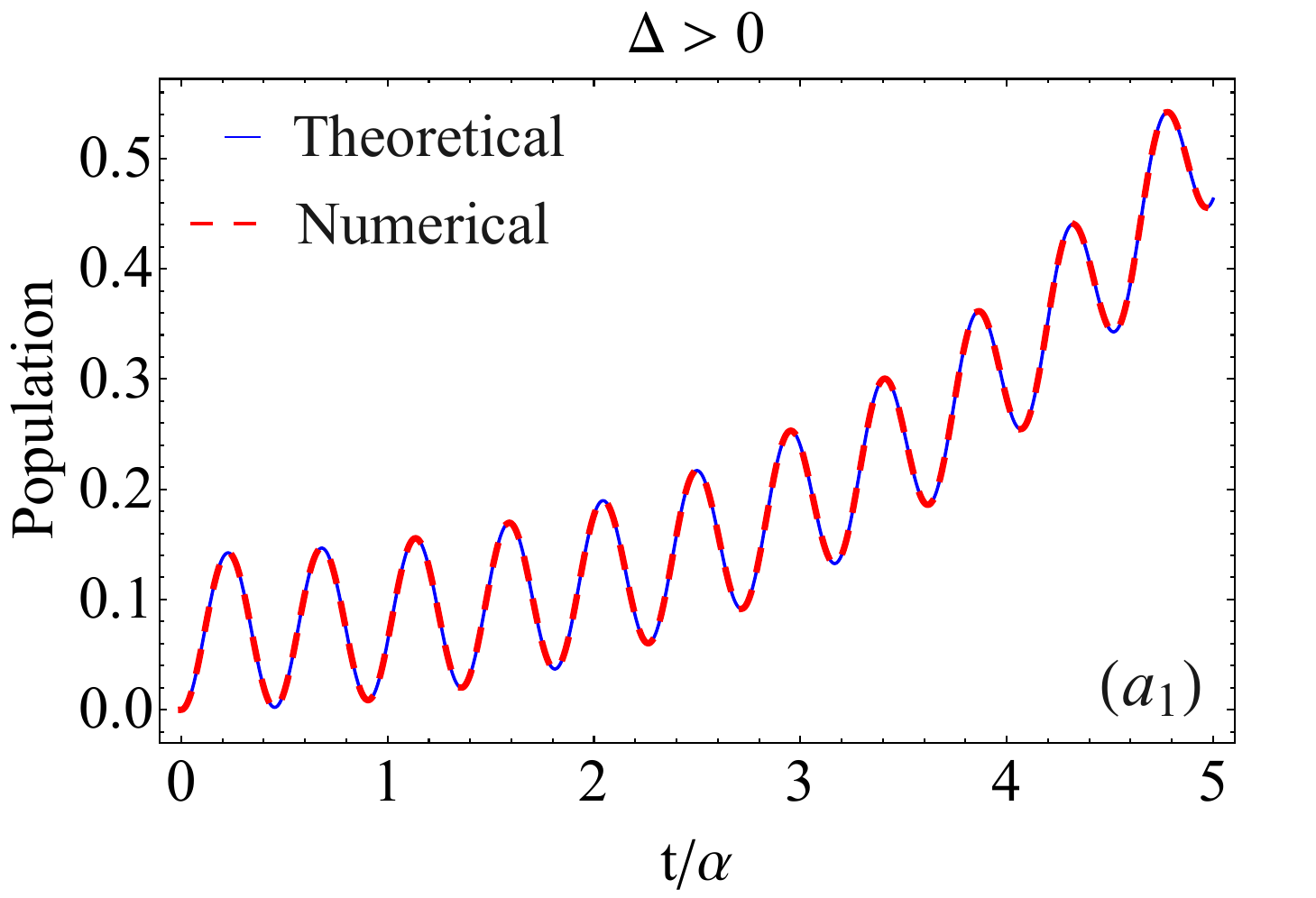}
	\includegraphics[width=0.23\textwidth, height =0.25\textwidth]{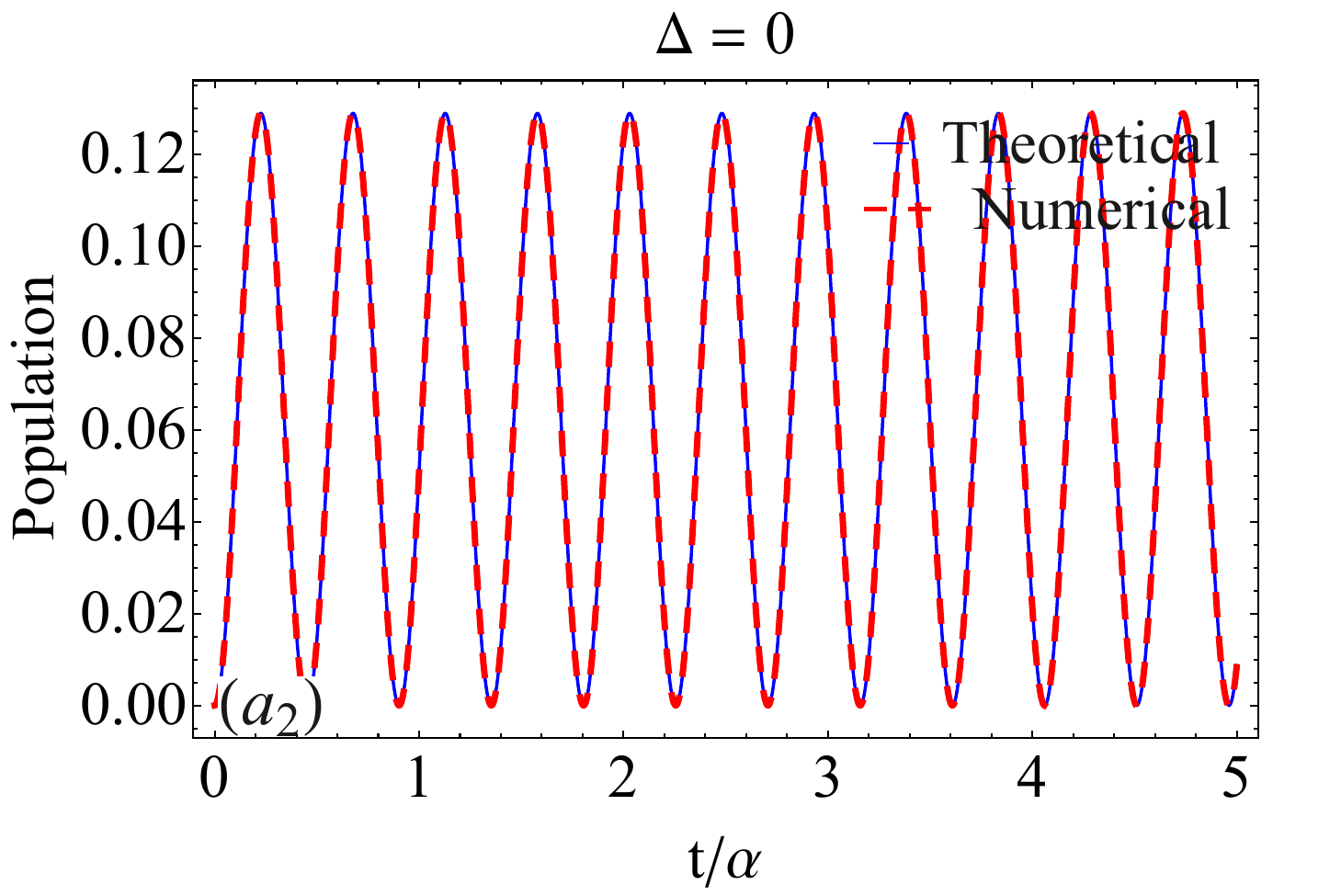}
	\includegraphics[width=0.23\textwidth, height =0.25\textwidth]{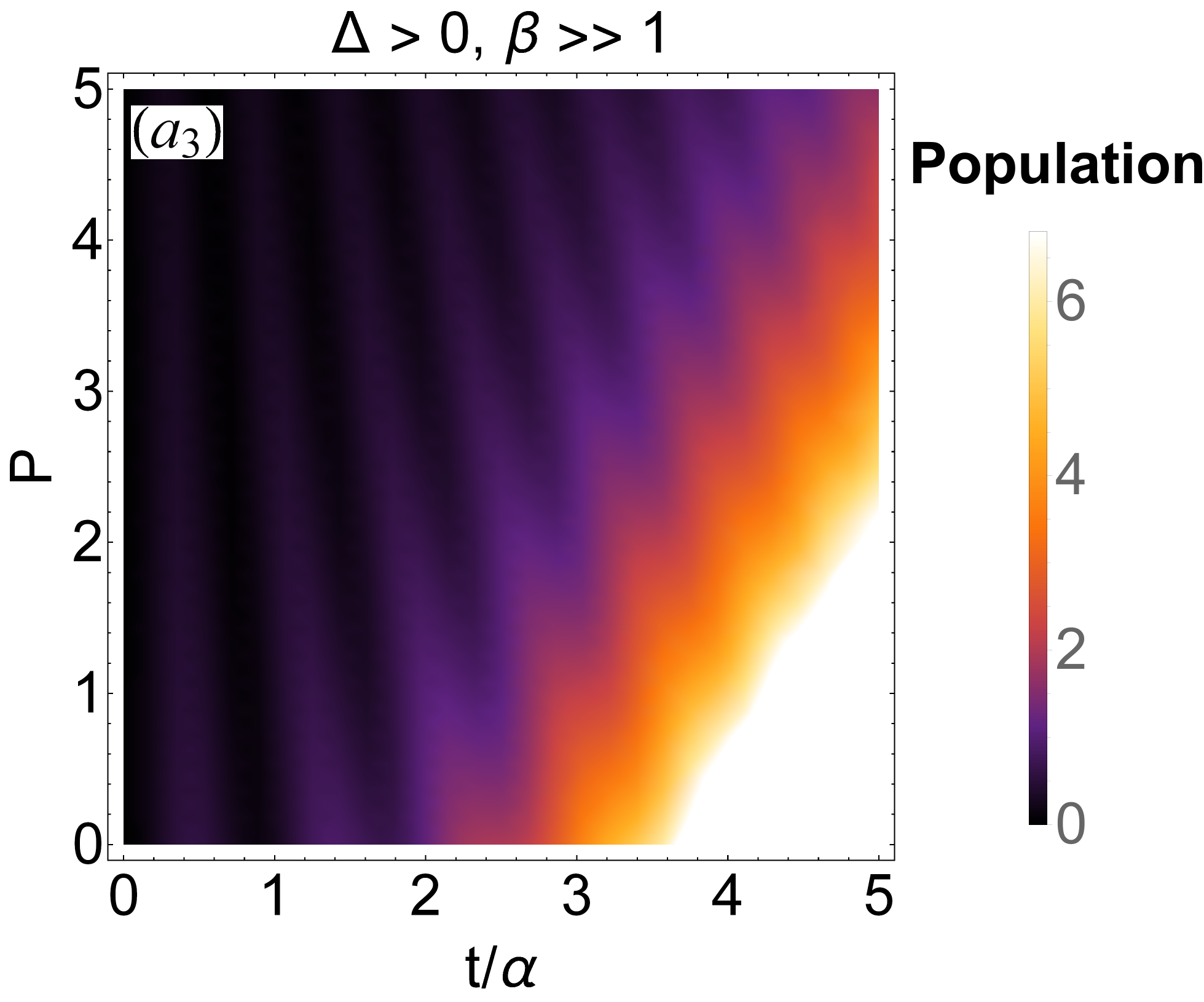}
	\includegraphics[width=0.23\textwidth, height =0.25\textwidth]{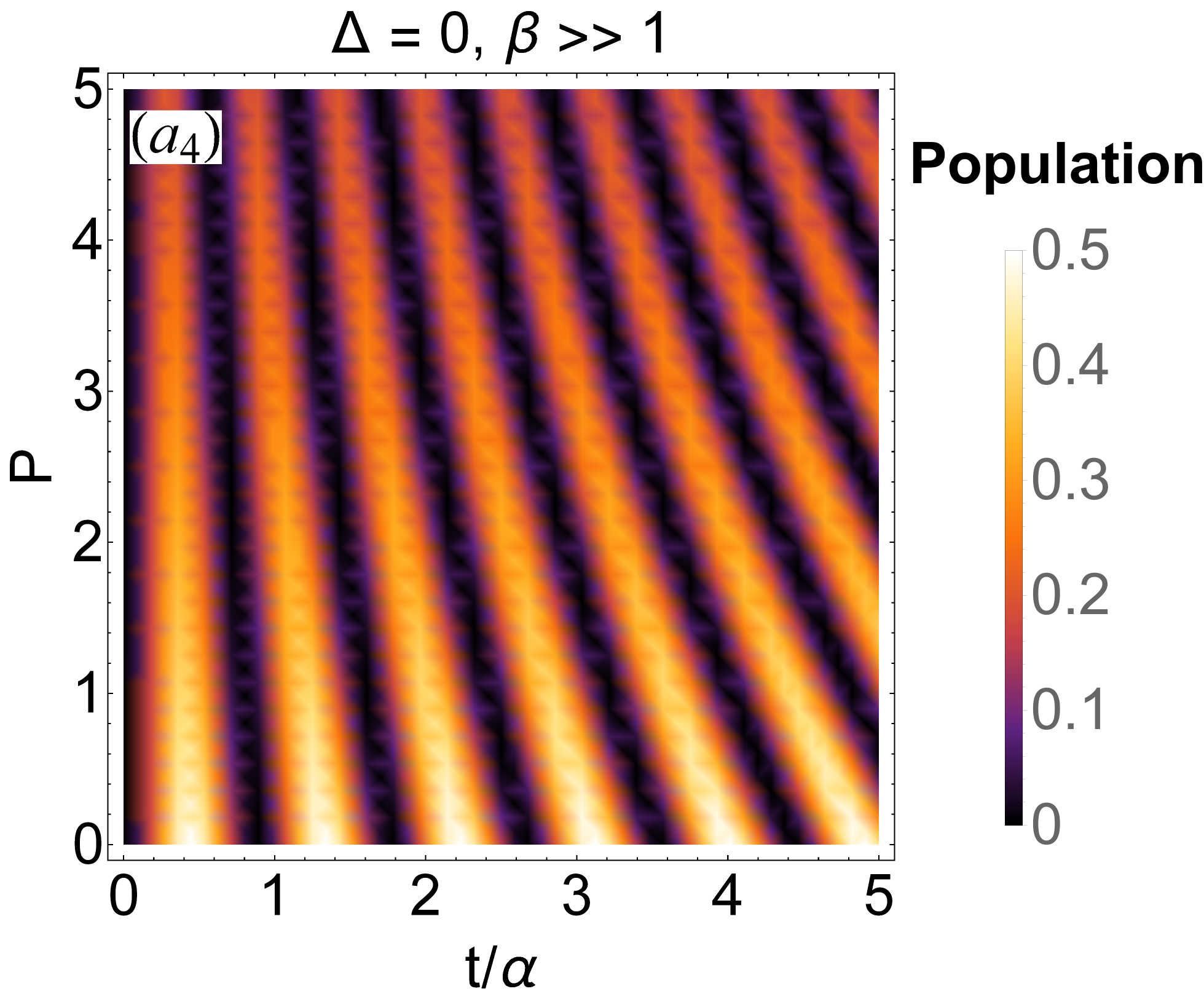}
	\caption{\small (color online) ($(a_1)$-$(a_2)$: Analogy between theoretical and numerical simulations. $(a_3)$-$(a_4)$: Interferogram of the population versus time as the coupling varies. We considered $P= 8.0/\alpha, \kappa = 5/\alpha$. The analytical results (blue line) agree with the numerical solutions (red line).}\label{fig2}
\end{figure}

with the help of the Mathematica program and the work of Bateman and Erderlyi \cite{HBateman}  we are considering the properties the hypergeometric functions:
\begin{equation}
	\frac{{\partial F\left( {\rho ,\omega ,\gamma ,x} \right)}}{{\partial x}} = \frac{{\rho \omega }}{\gamma }F\left( {\rho  + 1,\omega  + 1,\gamma  + 1,x} \right),\label{2.22a}
\end{equation}
\begin{equation}
	\begin{gathered}
		\frac{{\partial {x^{1 - \gamma }}F\left( {\rho  - \gamma  + 1,\omega  - \gamma  + 1,2 - \gamma ,x} \right)}}{{\partial x}} \hfill \\
		= \left( {1 - \gamma } \right){x^{ - \gamma }}F\left( {\rho  - \gamma  + 1,\omega  - \gamma  + 1,1 - \gamma ,x} \right) \hfill \\.\label{2.23}
	\end{gathered}
\end{equation}

\begin{figure}[h!]
	\centering
	\includegraphics[width=0.23\textwidth, height =0.25\textwidth]{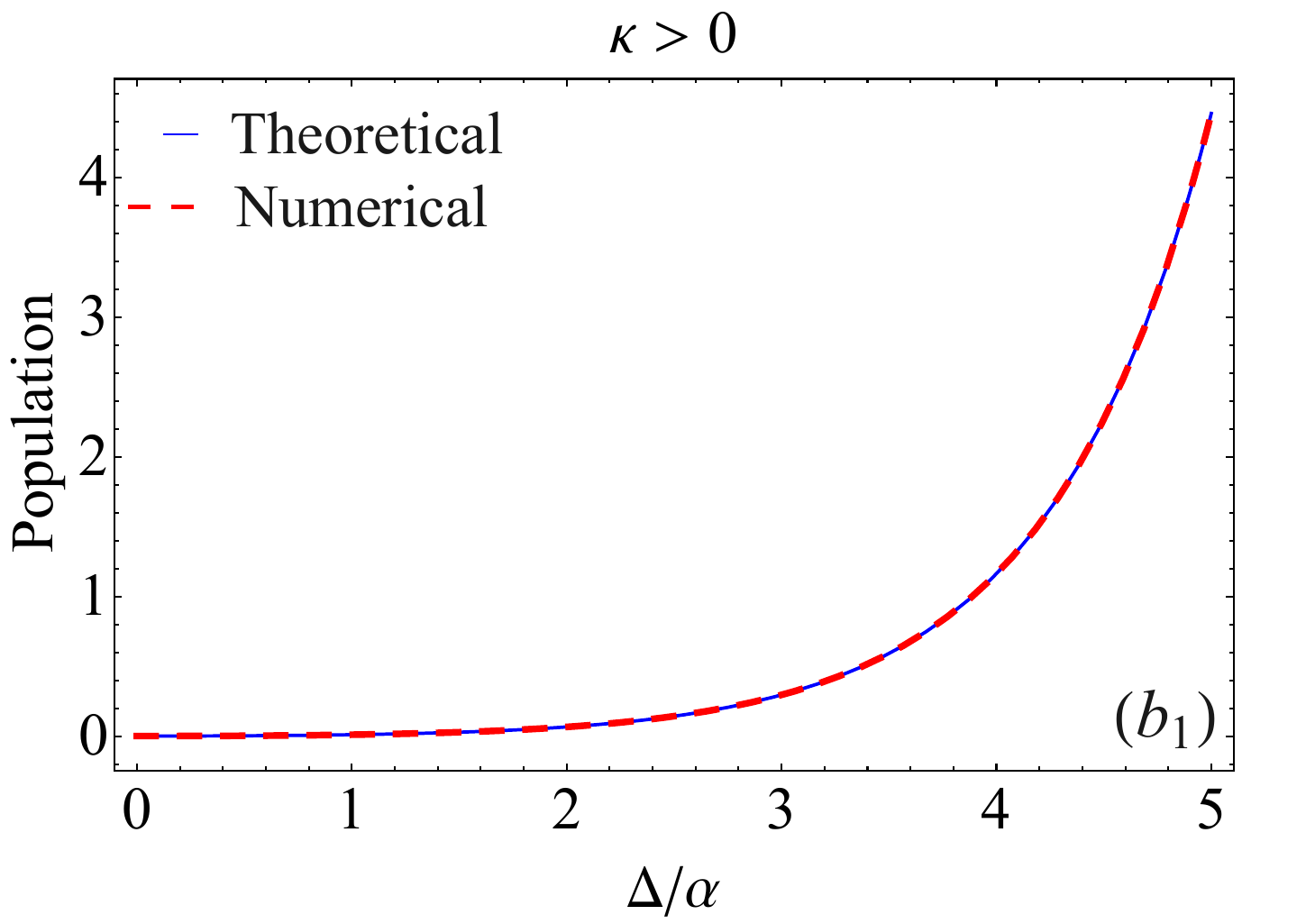}
	\includegraphics[width=0.23\textwidth, height =0.25\textwidth]{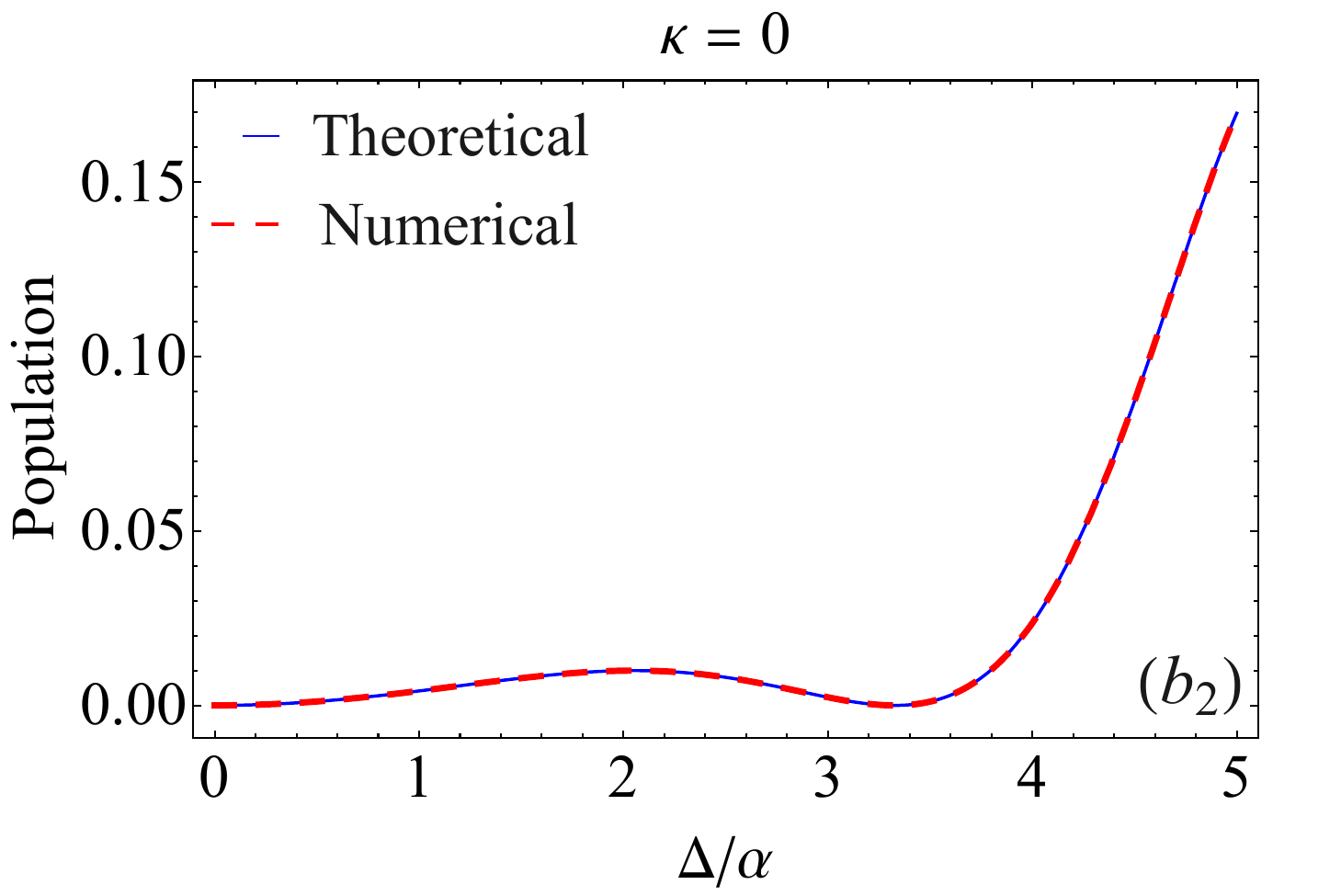}
	\includegraphics[width=0.23\textwidth, height =0.25\textwidth]{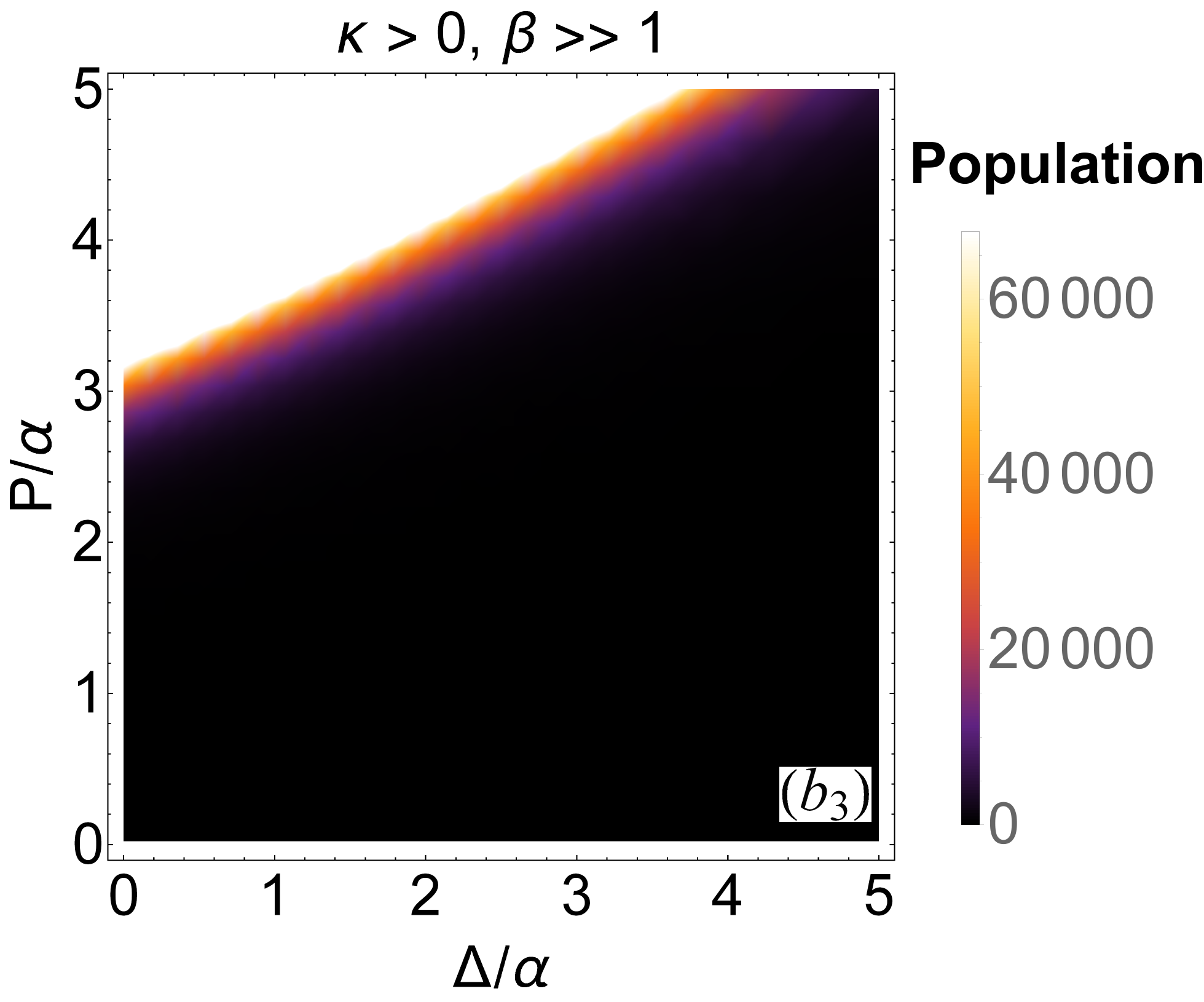}
	\includegraphics[width=0.23\textwidth, height =0.25\textwidth]{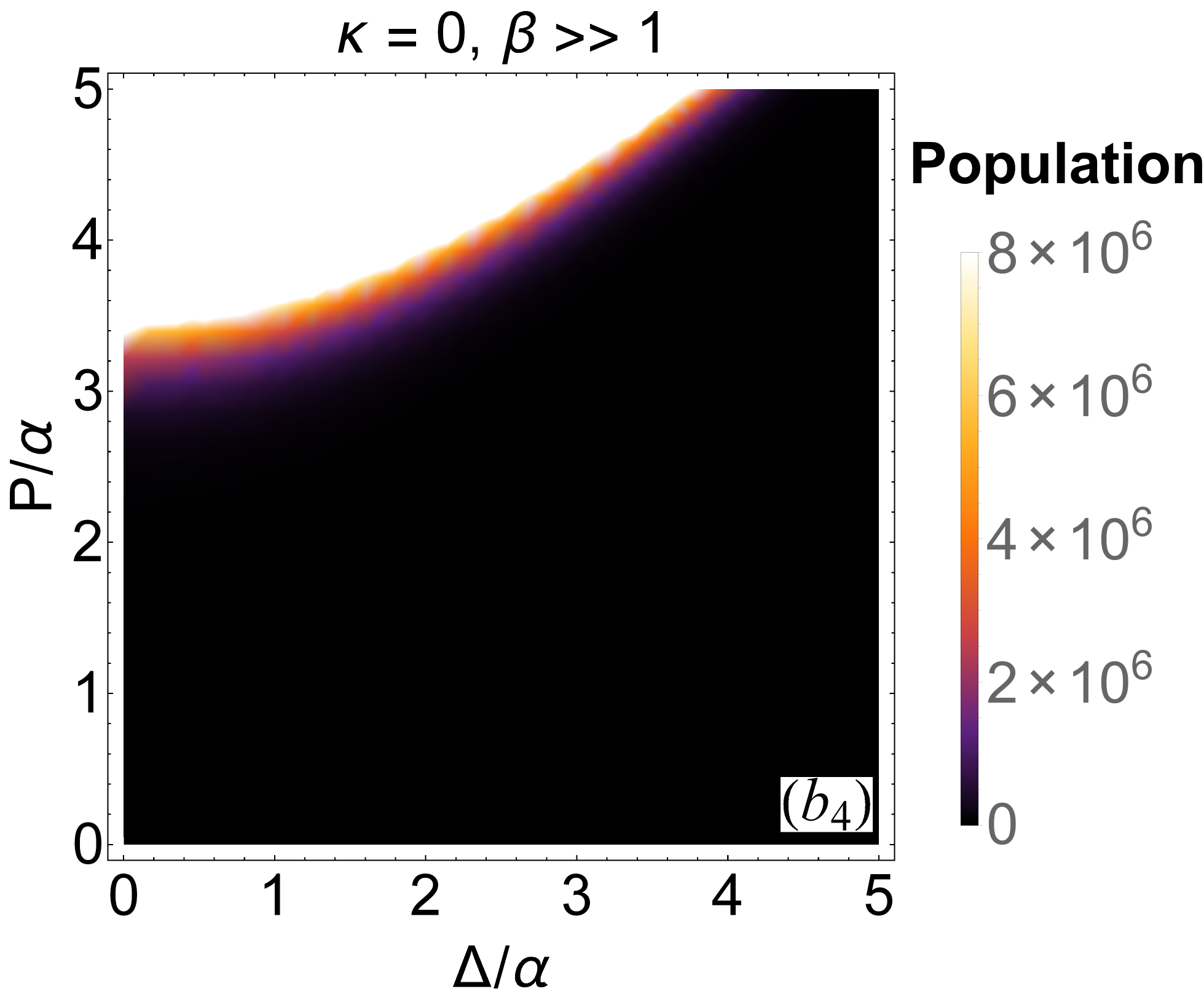}
	\caption{\small (color online) $(b_1)-(b_2)$: Analogy between theoretical and numerical simulations. $(b_3)-(b_4)$: Interferogram of the population against the coupling as the phase varies. The analytical results (blue line) perfectly agree with the numerical solutions (red dashed line).}\label{fig3}
\end{figure}

The constants ${b_ \pm }x_0^{ - \lambda }$ are determined using Ref\cite{Kenmoe}. To study the evolution of the system between the initial $t_0$ and an arbitrary time $t$. this propagator is determined by $B\left( x \right) = U\left( {x,{x_0}} \right)B\left( {{x_0}} \right)$. The components of the propagator are given by:
\begin{equation}
	{U_{\omega 1}}\left( {x,{x_0}} \right) = \frac{{{\Upsilon _{\omega 2}}\left( {x,{x_0}} \right)}}{{{\Upsilon _{12}}\left( {x,{x_0}} \right)}}\exp \left( {\chi \ln \frac{x}{{{x_0}}} + \varsigma \ln \frac{{1 - x}}{{1 - {x_0}}}} \right),
\end{equation}
\begin{equation}
	{U_{\omega 2}}\left( {x,{x_0}} \right) =  - \frac{{{\Upsilon _{\omega 1}}\left( {x,{x_0}} \right)}}{{{\Upsilon _{12}}\left( {x,{x_0}} \right)}}\exp \left( {\chi \ln \frac{x}{{{x_0}}} + \varsigma \ln \frac{{1 - x}}{{1 - {x_0}}}} \right).\label{2.21}
\end{equation}

The matrices elements are given by:
\begin{equation}
	{\Upsilon _{kk'}}\left( {x,{x_0}} \right) = {T_k}\left( {{x_0}} \right){R_{k'}}\left( x \right) - {R_k}\left( {{x_0}} \right){T_{k'}}\left( x \right)
\end{equation}
\begin{equation}
	{\Upsilon _{12}}\left( {{x_0},{x_0}} \right) = {R_1}\left( {{x_0}} \right){T_2}\left( {{x_0}} \right) - {R_2}\left( {{x_0}} \right){T_1}\left( {{x_0}} \right).
\end{equation}

Considering the wronskian linking the hypergeometric functions
\begin{equation}
	W = \left( {1 - \gamma } \right){x^{ - \gamma }}{\left( {1 - x} \right)^{\gamma  - \rho  - \omega  - 1}},\label{2.22}
\end{equation}

we easily obtain the parameter
\begin{equation}
	{\Upsilon _{12}}\left( {{x_0},{x_0}} \right) = \frac{i}{c}\left( {1 - \gamma } \right)x_0^{2\mu  - \gamma  + 1}{\left( {1 - {x_0}} \right)^{2\nu  + \gamma  - \rho  - \omega }}.
\end{equation}

Using the relation \eqref{2.23} the propagator element is expressed as:
\begin{equation}
	\begin{gathered}
		{U_{12}}\left( {x,{x_0}} \right) = ic\frac{{\Gamma \left( \rho  \right)\Gamma \left( \beta  \right)}}{{\Gamma \left( {\rho  + \beta  + 1 - \gamma } \right)\Gamma \left( \gamma  \right)}} \times  \hfill \\
		\left( {G_{\rho \beta }^\gamma \left( {x,{x_0}} \right) - G_{\rho \beta }^\gamma \left( {{x_0},x} \right)} \right)\exp \left( {\vartheta \left( {x,{x_0}} \right)} \right) \hfill \\ 
	\end{gathered}
\end{equation}
\begin{equation}
	\begin{gathered}
		{U_{11}}\left( {x,{x_0}} \right) =  - \frac{{\Gamma \left( \rho  \right)\Gamma \left( \beta  \right)}}{{\Gamma \left( {\rho  + \beta  + 1 - \gamma } \right)\Gamma \left( \gamma  \right)}} \times  \hfill \\
		{\left( {{x^\mu }{{\left( {1 - x} \right)}^\nu }G_{\rho \beta }^\gamma \left( {x,{x_0}} \right) - G_{\rho \beta }^\gamma \left( {{x_0},x} \right)} \right)^\prime }\exp \left( {\vartheta \left( {x,{x_0}} \right)} \right) \hfill \\ 
	\end{gathered},
\end{equation}

the survival probability is obtained by:
\begin{equation}
	{P_{22}}\left( {x,{x_0}} \right) = \operatorname{Re} \left( {{U_{22}}\left( {x,{x_0}} \right)} \right)^2 + \operatorname{Im} \left( {{U_{22}}\left( {x,{x_0}} \right)} \right)^2.
\end{equation}

\begin{figure}[h!]
	\centering
	\includegraphics[width=0.23\textwidth, height =0.25\textwidth]{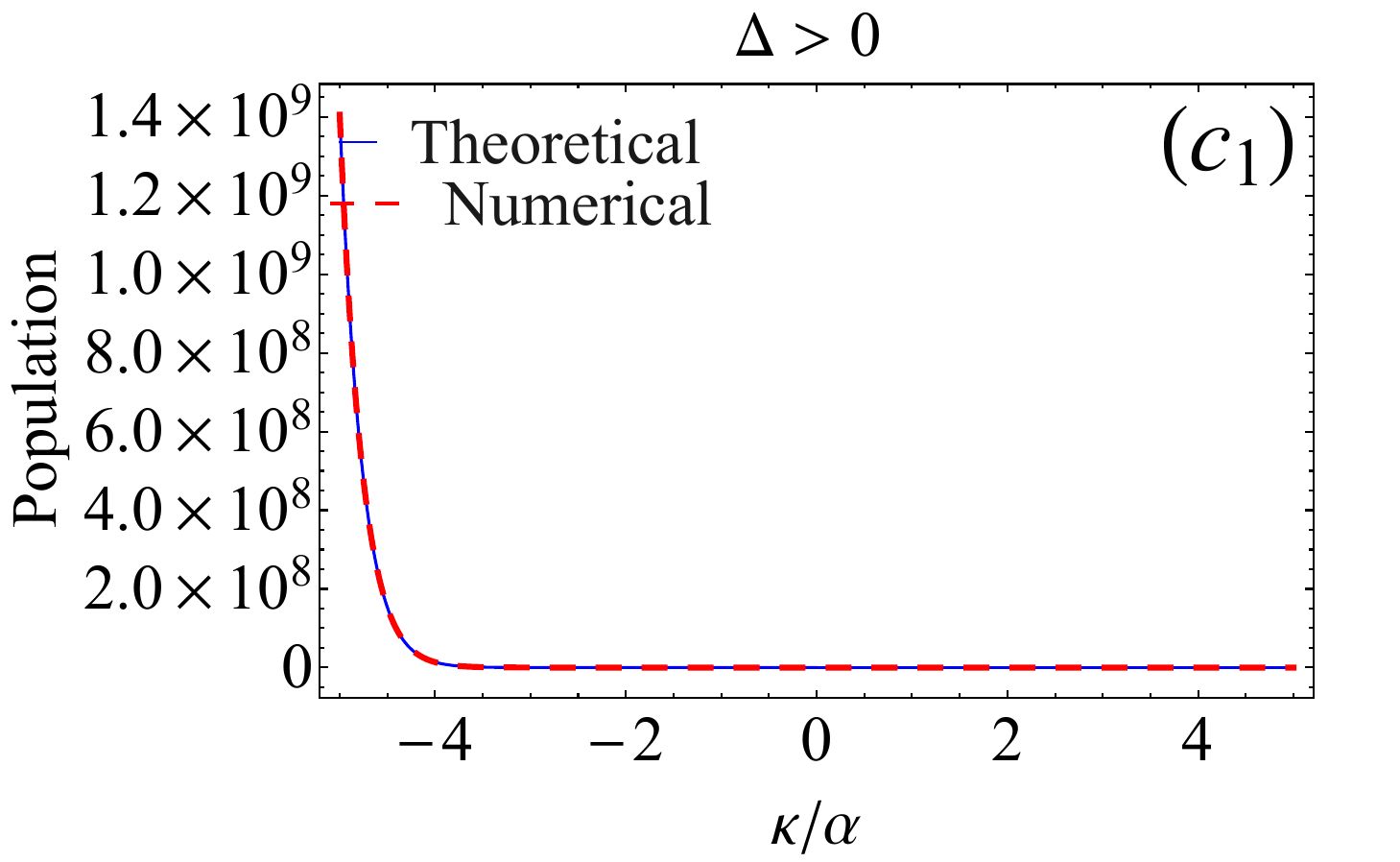}
	\includegraphics[width=0.23\textwidth, height =0.25\textwidth]{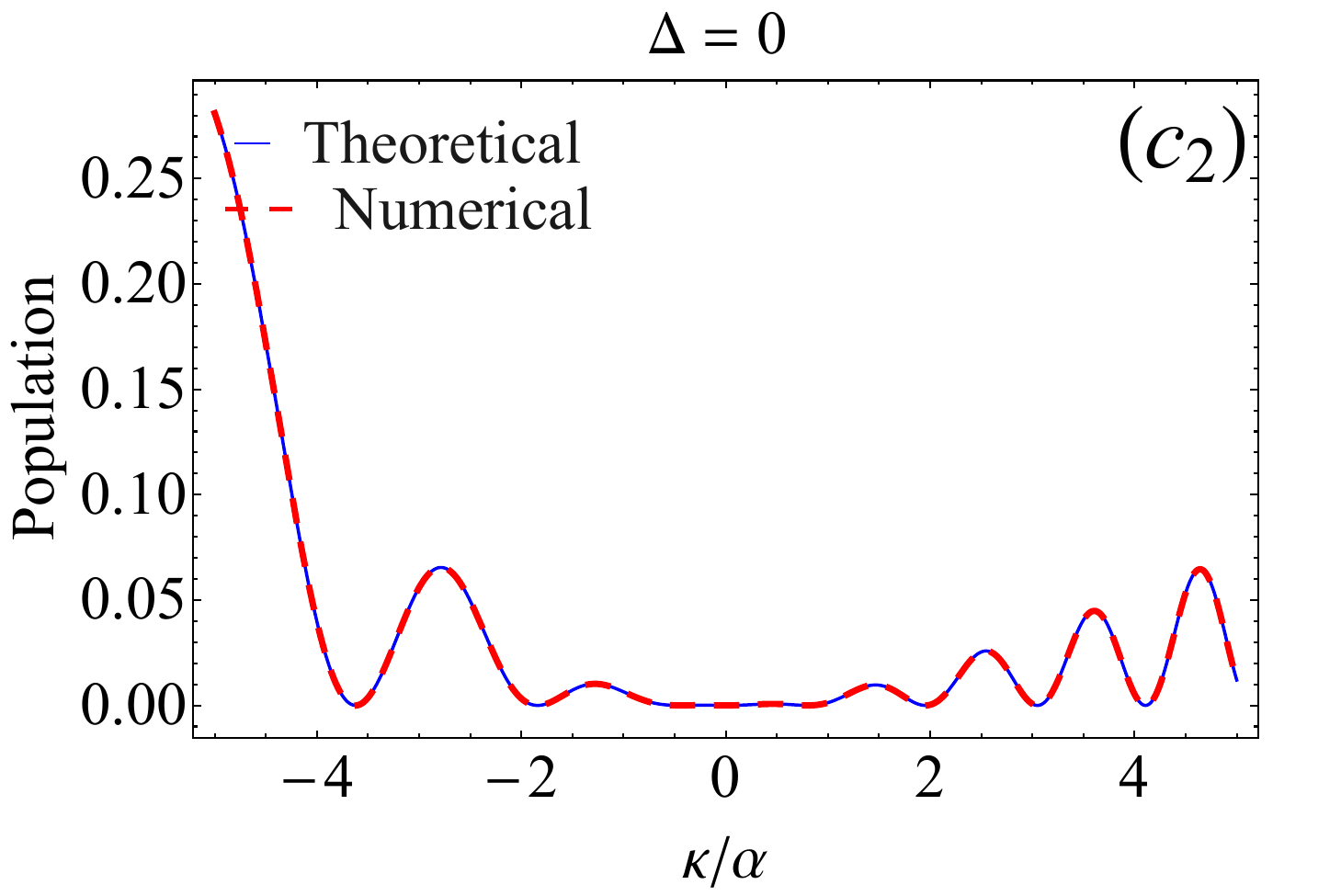}
	\includegraphics[width=0.23\textwidth, height =0.25\textwidth]{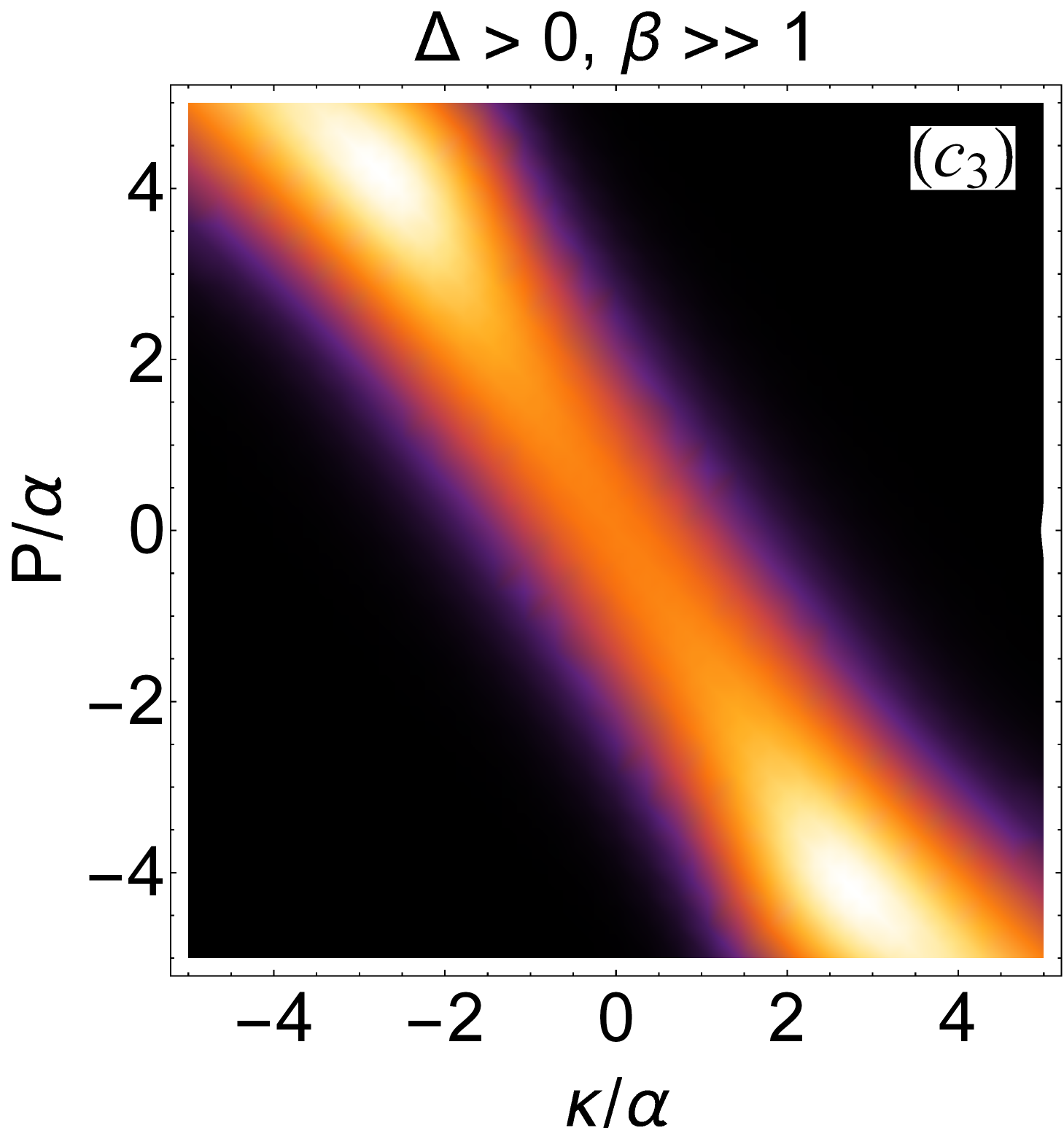}
	\includegraphics[width=0.23\textwidth, height =0.25\textwidth]{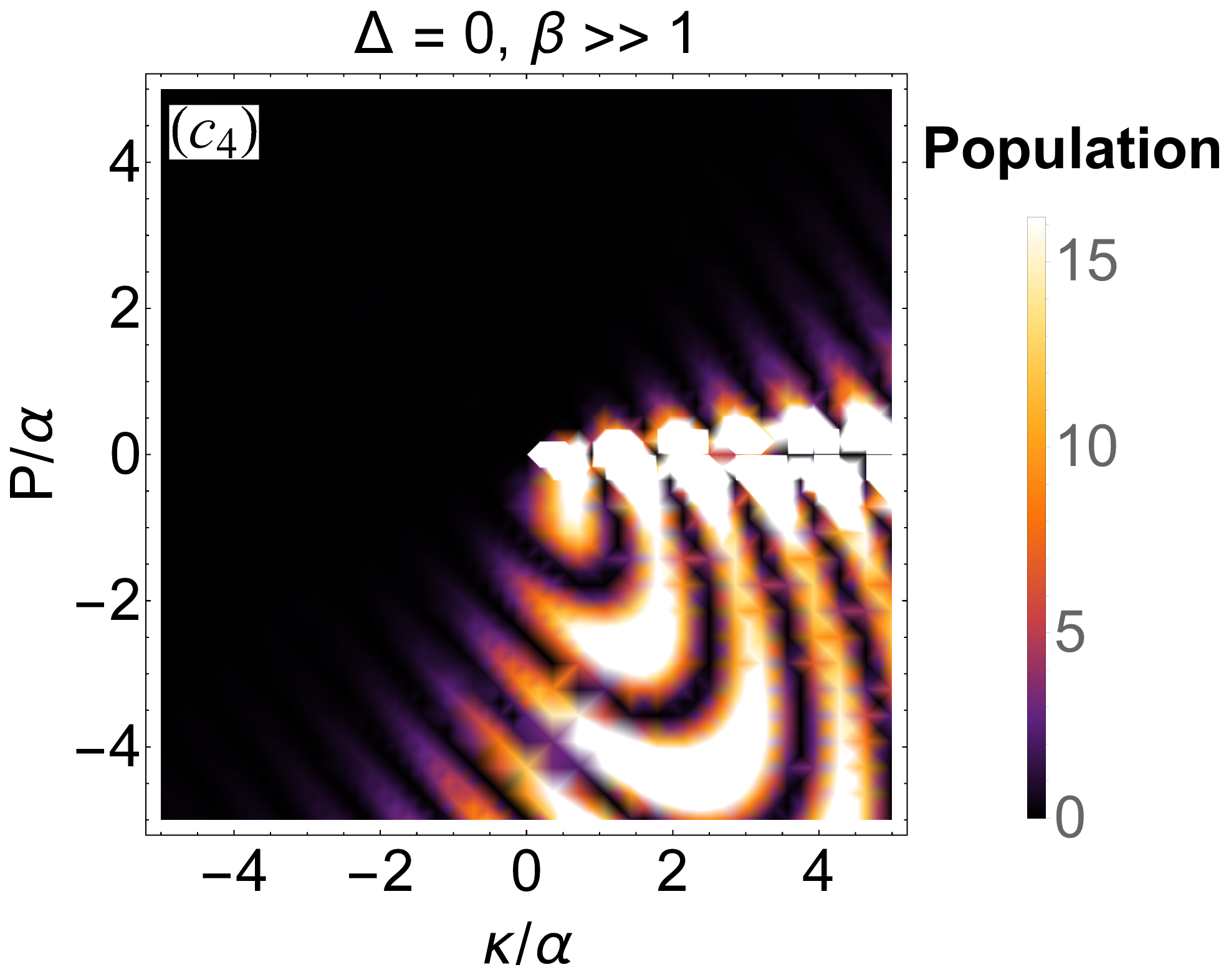}
	\caption{\small (Color online) $(c_1)-(c_2)$: Analogy between theoretical and numerical simulations. $(c_3)-(c_4)$: Variation of the population interferogram as a function of the shift. We have considered $A= 8.0/\alpha, B = 7.0/\alpha, \Delta = 1.0/\alpha $. The analytical results (blue line) perfectly agree with the numerical solutions (red dashed line).}\label{fig4}
\end{figure}

The probability of transition is given by:
\begin{equation}
	{P_{12}}\left( {x,{x_0}} \right) = \operatorname{Re} \left( {{U_{12}}\left( {x,{x_0}} \right)} \right) + \operatorname{Im} \left( {{U_{12}}\left( {x,{x_0}} \right)} \right).
\end{equation}

\subsection{Discussions of our results}

Our principal results are presented in Figs.\ref{fig2}, \ref{fig3}, and \ref{fig4}  where we plot the variation of the populations and their interferograms in different cases. It is important to note that we do not evaluate the cases where $\Delta < 0$ and $\kappa < 0$,  because they provide similar information to the case $\Delta > 0$ and $\kappa > 0$.

In Fig.\ref{fig2}, we highlight the variation of the population as a function of time. In frame ($a_1$) when $\Delta > 0$,  the population is described by an oscillation with a low amplitude, which explains why the interferogram of this population shows a light beam with successive transparent dark and shiny sections. However, when $\Delta = 0$, the oscillations of the population are more pronounced in ($a_2$), which is reflected by visible light beams with a succession of dark and shiny sections.

Fig.\ref{fig3} illustrates the evolution of the population versus the coupling $\Delta/\alpha$. The population in frame ($b_1$) increases and exceeds 1 due to the high values of the shift. Its interferogram in ($b_3$) shows the same behavior, with a light beam becoming more condensed, featuring a large dark part and a short shiny section. In frames ($b_2$) and ($b_4$), we observe the same phenomenon described in ($b_1$) and ($b_3$), with the presence of slow oscillations.

In Fig.\ref{fig4} we present the variation of the population of the tanh model against the shift $\kappa/\alpha$ when the coupling $\Delta$ varies. In frame ($c_1$),  the population exceeds 1 due to the high values of the coupling, which amplify the amplitude of the population. Its interferogram, with $\beta  \gg 1 $, shows the interaction between condensed light beams, which is very useful in interferometry.
In other cases, when imaginary dissapears in ($c_2$), the population describes oscillations, but the interesting phenomenon appears in its interferogram in ($c_4$), with the presence of light beams in the form of a spiral with dark and shiny sections. This represents the interferometry of our model.

\section{Energies of the tanh model}\label{sec5}

To study the energy of the tanh model, we examine the eigenvalues of our system. This enables us to obtain:
\begin{equation}
	{E_ \pm } =  \pm \Theta \csc {2\vartheta \left( t \right)},
\end{equation}
where
\begin{equation}
	\tan {2\vartheta \left( t \right)} = \frac{{\Theta }}{{\Xi \left( t \right)}}.
\end{equation}

\begin{figure}[h!]
	\centering
	\includegraphics[width=0.23\textwidth, height =0.20\textwidth]{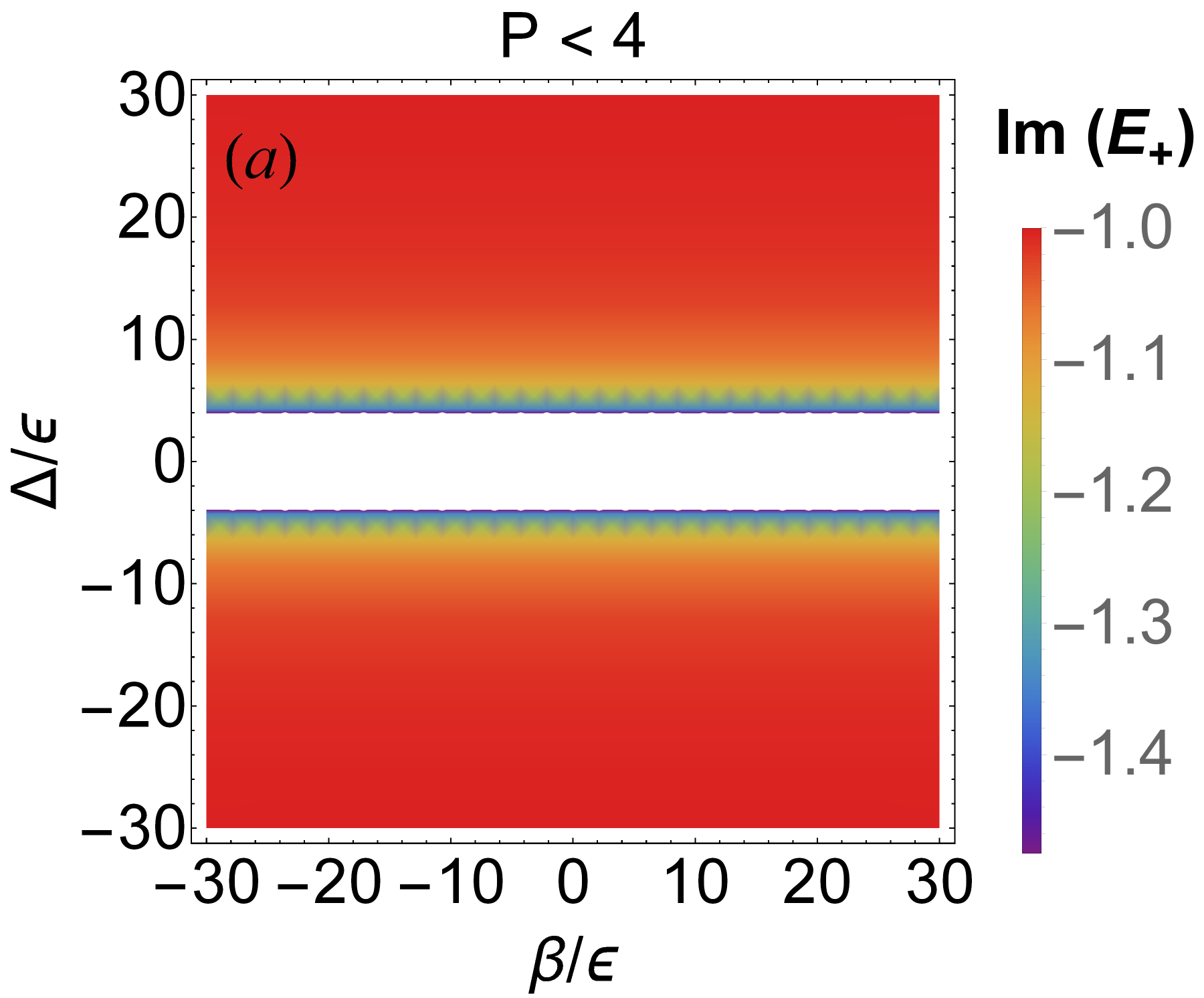}
	\includegraphics[width=0.23\textwidth, height =0.20\textwidth]{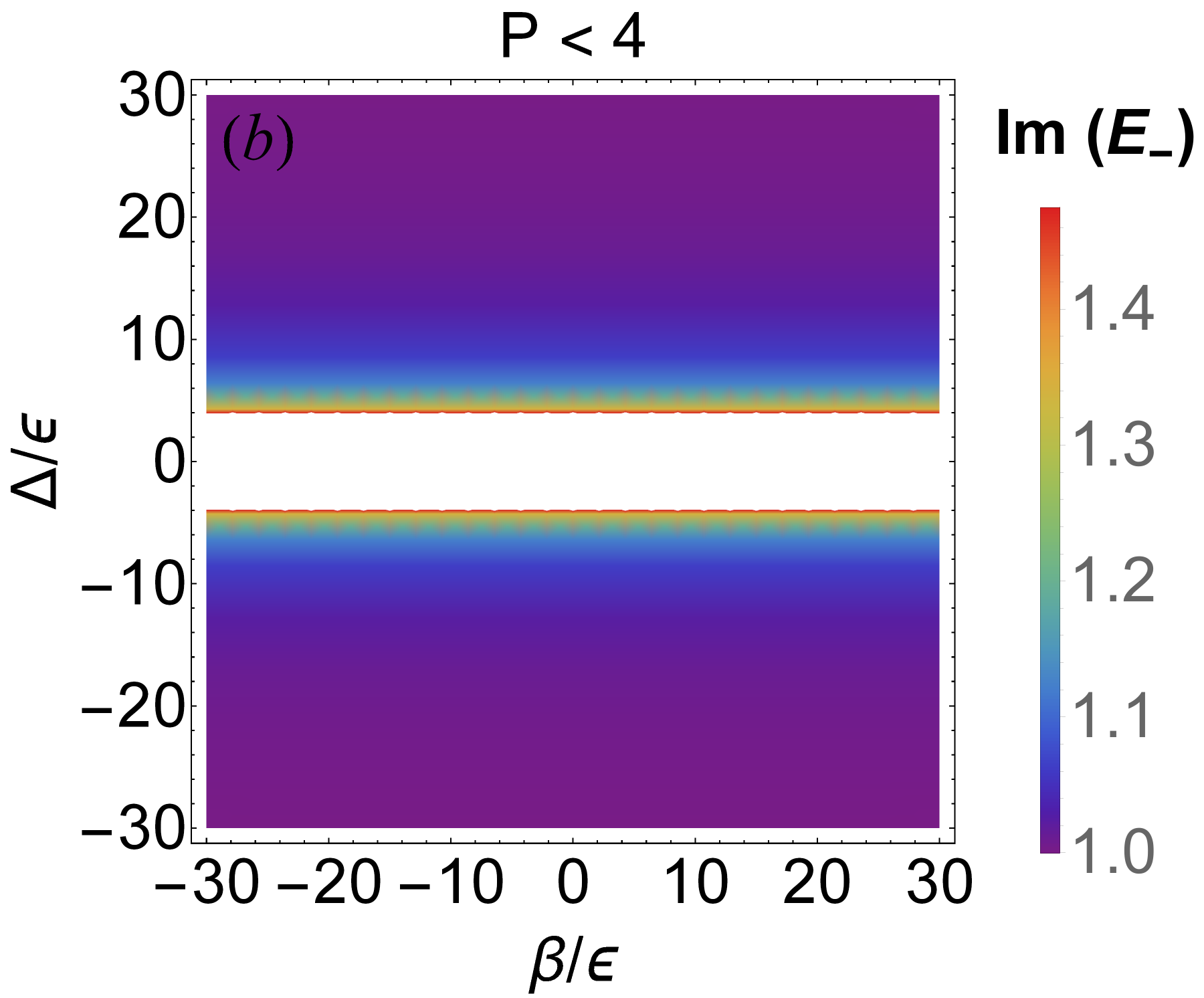}
	\includegraphics[width=0.23\textwidth, height =0.20\textwidth]{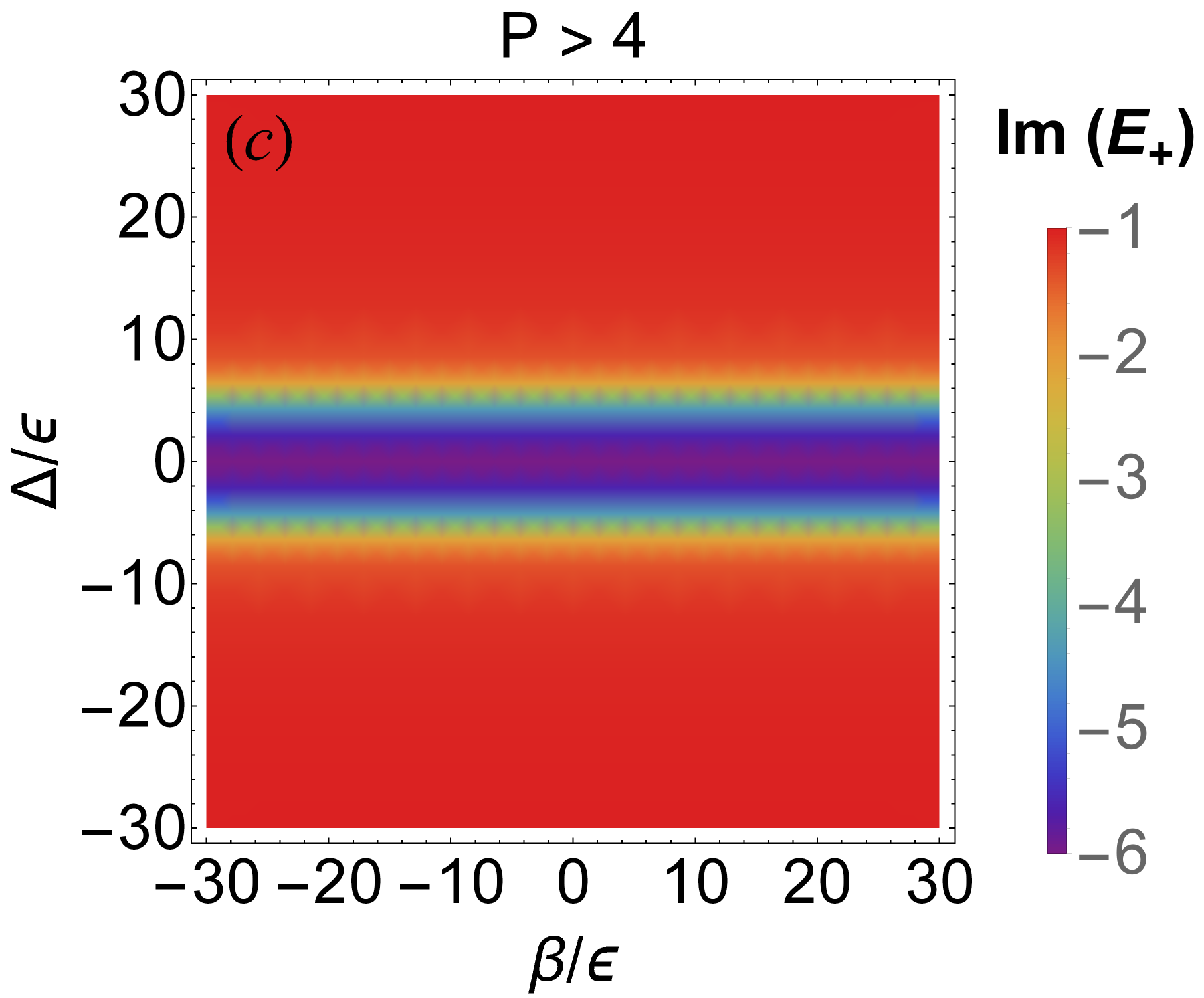}
	\includegraphics[width=0.23\textwidth, height =0.20\textwidth]{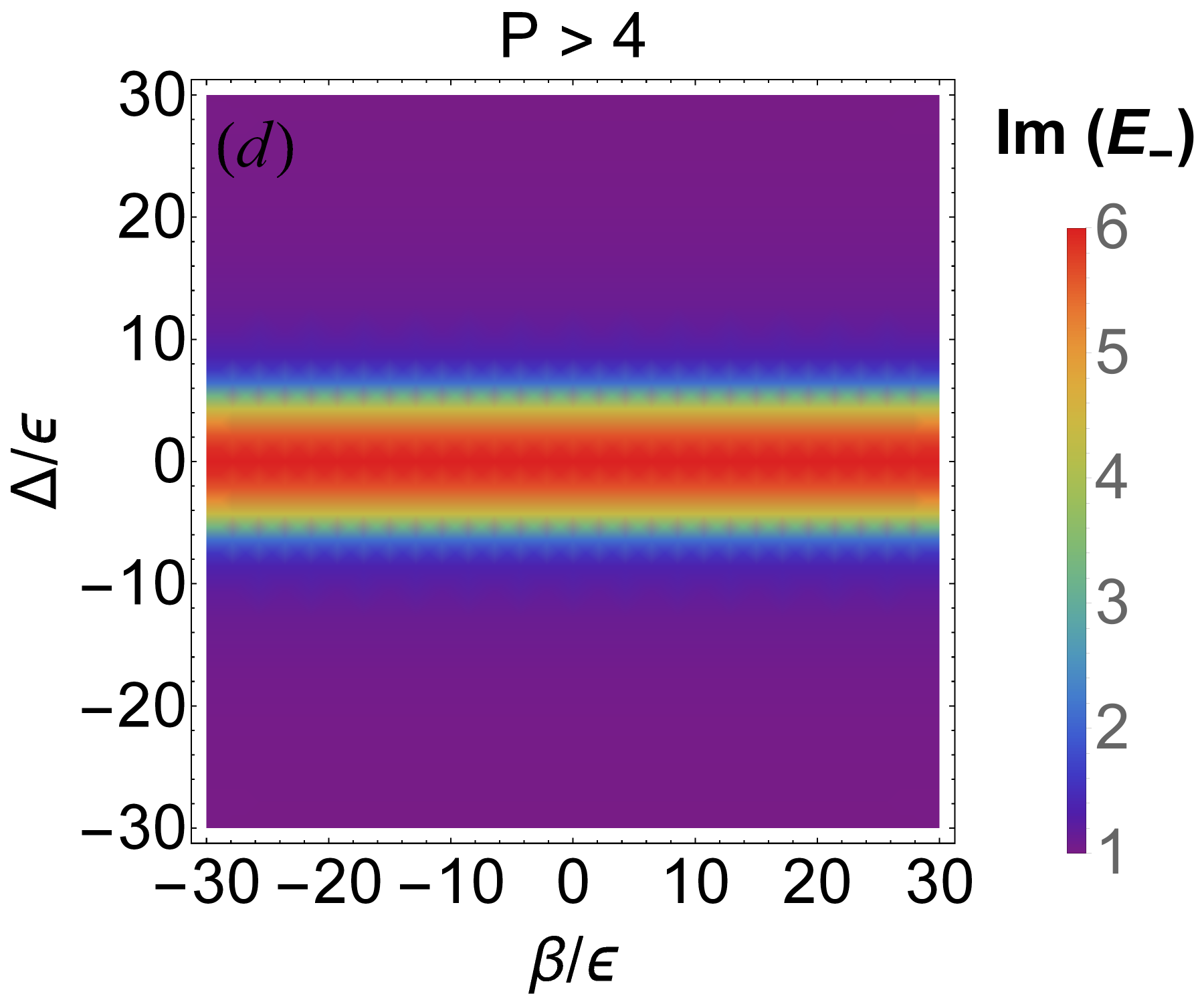}
	\caption{\small (Color online) Plots of the real part of the energy against the coupling and the phase as the amplitude varies. In this figure, we have considered $t = 5/\epsilon, \alpha = 20/\epsilon $. This figure shows the allowed and forbidden zone for the transmission of energy, this kind of interferogram can be the signature of the conduction in some cases. In this figure, we use the rainbow of Mathematica.}\label{fig5}
\end{figure}

We note that this eigen-energy is complex, with the contribution of an imaginary coupling term. This also explains the non-Hermitian character of this model. We can separate it into real and imaginary terms.

\begin{equation}
	\operatorname{Re} \left( E_\pm \right) = \frac{1}{2}\left| {{Z^{1/2}}\left( t \right)} \right|\cos \phi,
\end{equation}
\begin{equation}
	\operatorname{Im} \left( E_\pm \right) = \frac{1}{2}\left| {{Z^{1/2}}\left( t \right)} \right|sin\phi,
\end{equation}

\begin{figure}[h!]
	\centering
	\includegraphics[width=0.23\textwidth, height =0.20\textwidth]{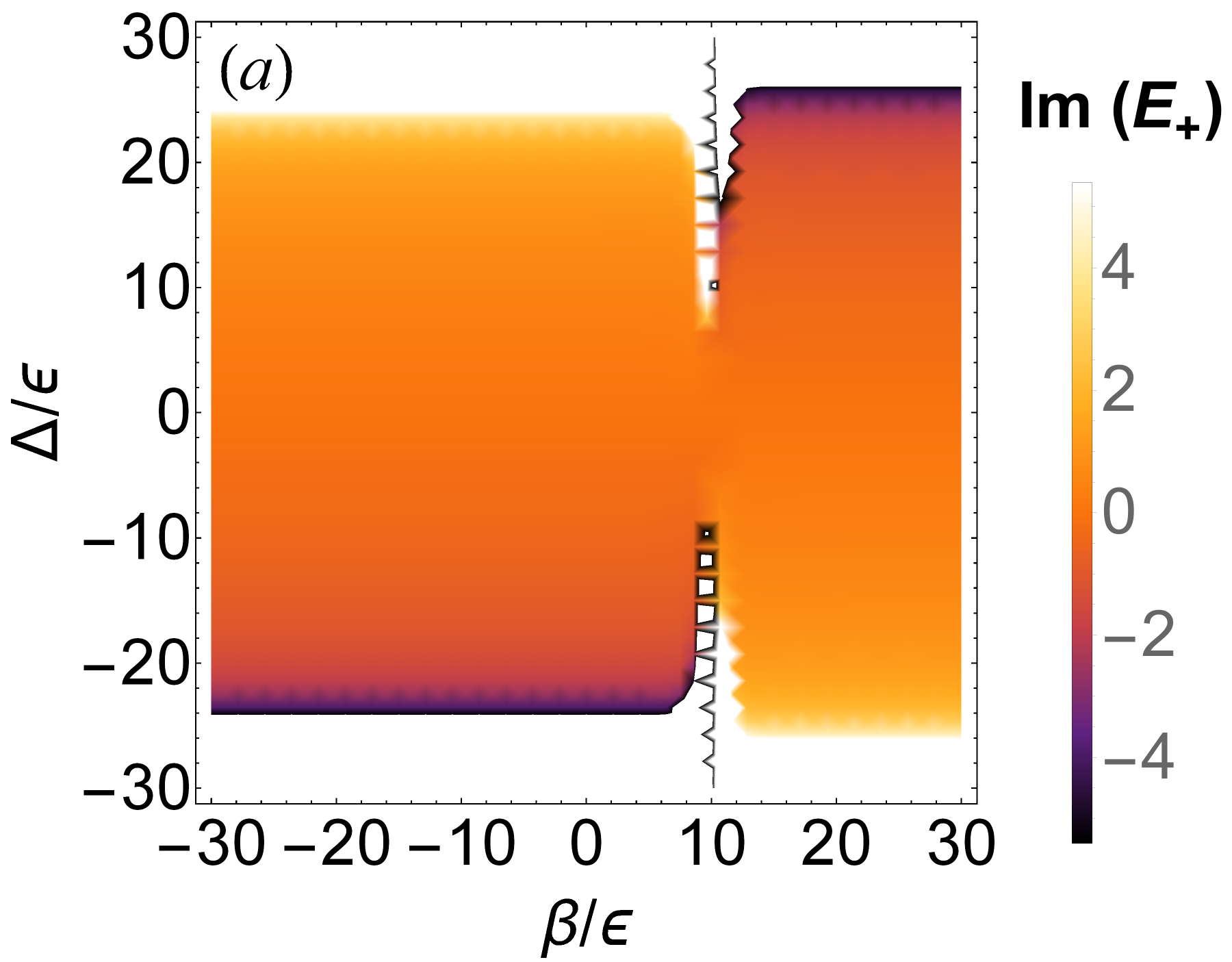}
	\includegraphics[width=0.23\textwidth, height =0.20\textwidth]{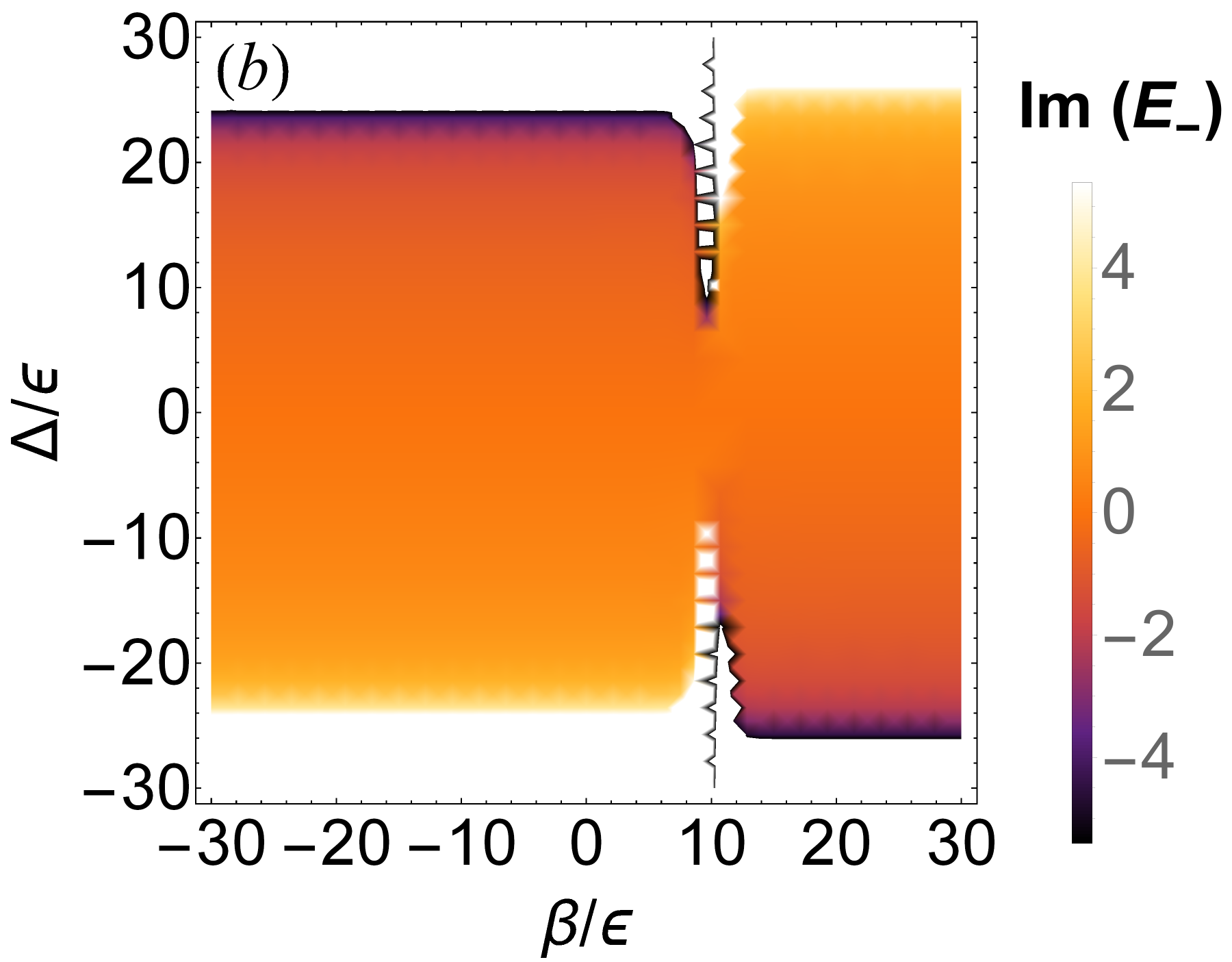}
	\caption{\small (Color online) Plots of the imaginary part of the energy versus coupling and phase. In this figure we have considered $P= 25/\epsilon, t = -0.5/\epsilon, \alpha = 20/\epsilon $. The sweep speed $\alpha$ and the time $t$ are responsible for the order or chaos in the system for high values. Remember that the imaginary part of the energy describes the loss of quantum information. In this figure, we use the sunset color of Mathematica.}\label{fig6}
\end{figure}

 with
\begin{equation}
 	\tan \left( {2\phi } \right) = \frac{{2\kappa \Delta }}{{{\Omega ^2}\left( t \right) + {\kappa ^2} - {\Delta ^2}\left( t \right)}},
\end{equation}
\begin{equation}
	{Z^{1/2}}\left( t \right) = \frac{{2\kappa \Delta }}{{{\Omega ^2}\left( t \right) + {\kappa ^2} - {\Delta ^2}\left( t \right)}}.
\end{equation}

The diagrams in Fig.\ref{fig5} and Fig.\ref{fig6} show the evolution of the real and imaginary parts of the energy.

The real part of the energy shows the creation of the allowed and forbidden zones. The allowed zone corresponds to the area where communication between the ground and excited states is possible, while the forbidden area corresponds to the zone where communication between these states is not possible \cite{Kammogne}. We established the condition under which the forbidden zone appears, and we note that when $P < 4$, there is no communication in the system, as we can see in frames ($a$) and ($b$). In contrast, when $P > 4$, the two states can communicate, as shown in frames ($c$) and ($d$), allowing the transmission of quantum information. We can conclude that the amplitude plays a crucial role in determining the forbidden and allowed zones in the energy diagram.

The imaginary part of the energy in Fig. \ref{fig6} is responsible for energy loss. In this graph, we note that the time and sweep velocity influence the dynamics in the energy diagram, dictating the chaos and the order or disorder in the system. Our model presents similar behavior to other standard models, and we will develop it further in the next part of this work.

\section{Analogy with the rabi and Landau-Zener model}\label{sec6}

The first model of Demkov-Kunike presents the peculiarity of being close to both the Rabi and Landau-Zener models. To achieve this equivalence, on one hand, we consider the rapid time approximation with the help of a Maclaurin expansion. When the time goes to infinity, we obtain the Landau-Zener model, while in the opposite case, in the short-time approximation (when the time goes to zero), we have the Rabi model.

\subsection{Rabi model}

The Rabi model is obtained in the case of the short-time approximation when the time is zero and $\beta =0$. At this point, the detuning becomes $\Omega \left( t \right) = \kappa $.  We follow the same strategy described by \cite{Kammogne1} to obtain the transition and survival probabilities:
\begin{equation}
	{P_{R{a_2}}}\left( t \right) = \frac{1}{{{{\csc }^2}2\vartheta }}{\sin ^2}\left( {\frac{{ {\Theta } \csc 2\vartheta }}{2}t} \right),
\end{equation}

\begin{equation}
	{P_{Ra_1}}\left( t \right) = 1 - {P_{R{a_2}}}\left( t \right),\label{5.8}
\end{equation}

\begin{figure}[h!]
	\centering
	\includegraphics[width=0.23\textwidth, height =0.20\textwidth]{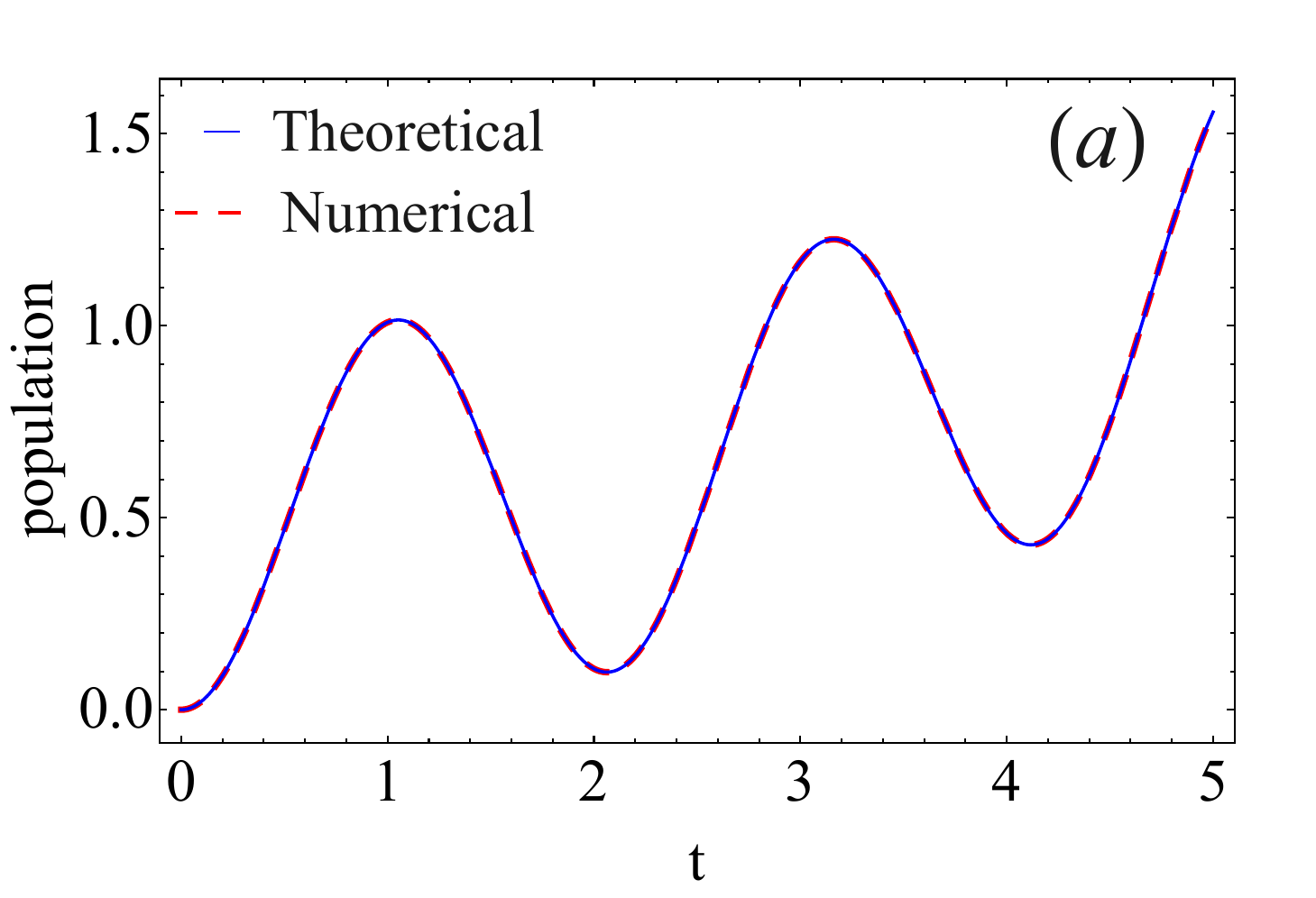}
	\includegraphics[width=0.23\textwidth, height =0.20\textwidth]{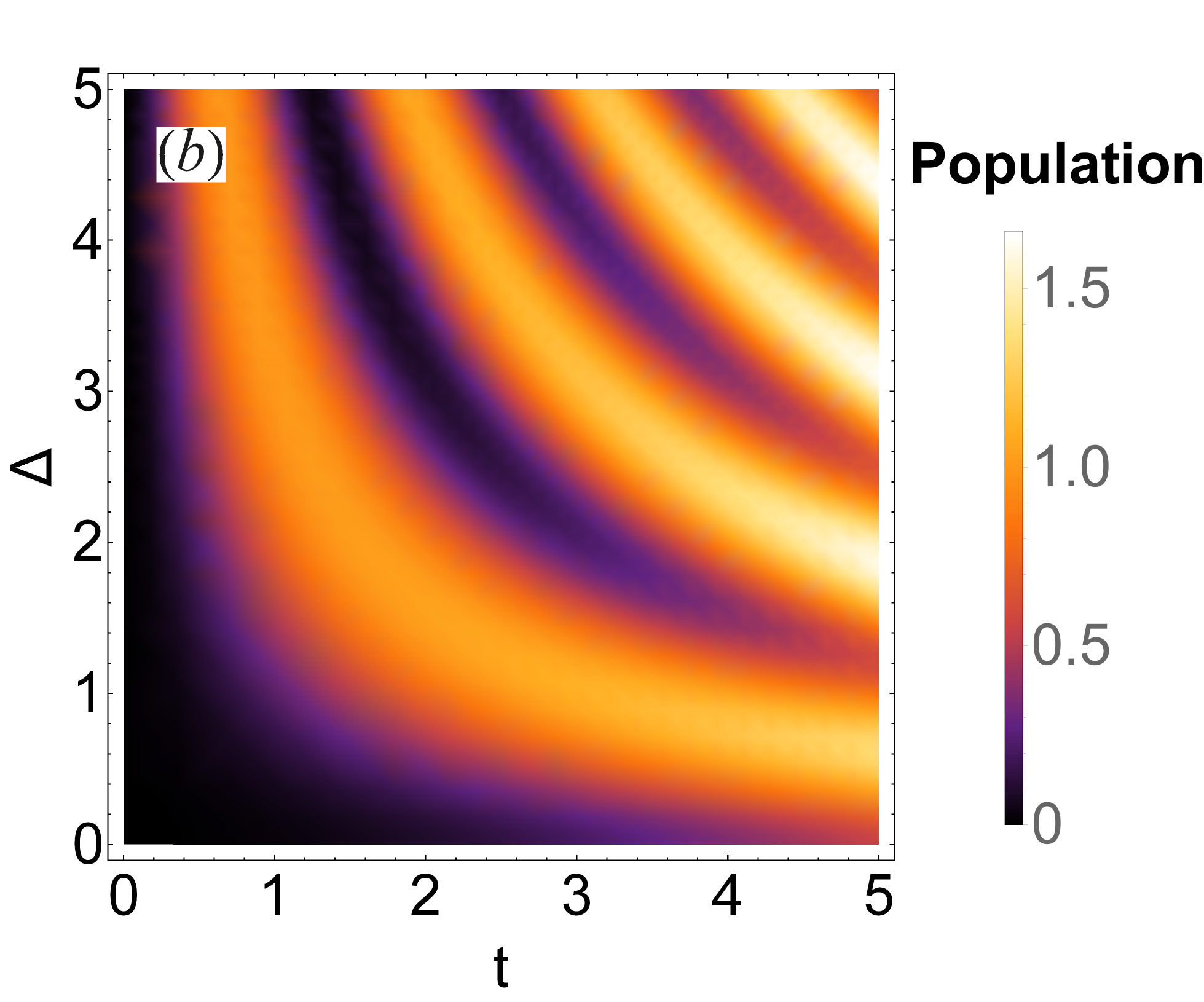}
	\caption{\small (Color online) (a) Similarity between theoretical and numerical simulations of the Rabi model with imaginary coupling. (b) Interferogram of the Rabi model with imaginary coupling. In this figure, we have considered $\kappa = 0.3/, \Delta > 0 $. Figure (b) shows how we can materialize the dark and light fringes using the jump $\kappa$. This alternate would be at the origin of the Rabi interferometer. Our theoretical results are similar to the numerical simulations.}\label{fig7}
\end{figure}

where the imaginary-dependent angle $\vartheta$ is defined by:

\begin{equation}
	\tan 2\vartheta  = \frac{{\Theta }}{\kappa }.
\end{equation}

Fig.\ref{fig7} shows the similarities between the theory and numerical data in (a),and the interferogram of the population as a function of time and coupling in (b). In (b), we observe a beam of light with bright and dark parts, propagating like a wave on the sea. In Fig.\ref{fig7} (a), we show that our analytical results agree with the numerical Rabi model, which is greater than 1 due to the imaginary nature of the coupling. In the next section, we present the analogy between our model and the Landau-Zener model.

\subsection{Landau-Zener model}

In this case, as we mentioned before, the most interesting approximation is the rapid time approximation when the time goes to infinity. The detuning takes the following form: $\Omega \left( t \right) = \alpha Pt + \kappa$ at the first order of the Maclaurin expansion. Using this configuration with the time-dependent Schr\"odinger equation, and with the help of the gauge transformation, we arrive at the second-order differential equation:
\begin{equation}
	\frac{{{d^2}{\psi _1}\left( t \right)}}{{d{t^2}}} \pm i\left( {\alpha Pt + \kappa } \right)\frac{{d{\psi _1}\left( t \right)}}{{dt}} + \frac{1}{4}{\left( {i\Delta  + \kappa } \right)^2}{\psi _1}\left( t \right) = 0. \label{5.1}
\end{equation}

To solve \eqref{5.1}, we use the ansatz transformation $x = \sqrt {\alpha P} \left( {t - \frac{\kappa }{{\alpha P}}} \right){e^{i\pi /4}}$, , followed by the change of variables ${\psi _{1,2}}\left( x \right) = {X_{1,2}}\left( x \right){e^{ \mp {z^2}/4}}$. Then, we can find the probability amplitudes \cite{KenmoeMB} as:
\begin{equation}
	\frac{{{d^2}{X_{1,2}}\left( x \right)}}{{d{t^2}}} + \left( {i\lambda  \mp 1/2 - {z^2}/4} \right){X_{1,2}}\left( x \right) = 0,\label{5.2}
\end{equation}

\begin{figure}[h!]
	\centering
	\includegraphics[width=0.23\textwidth, height =0.20\textwidth]{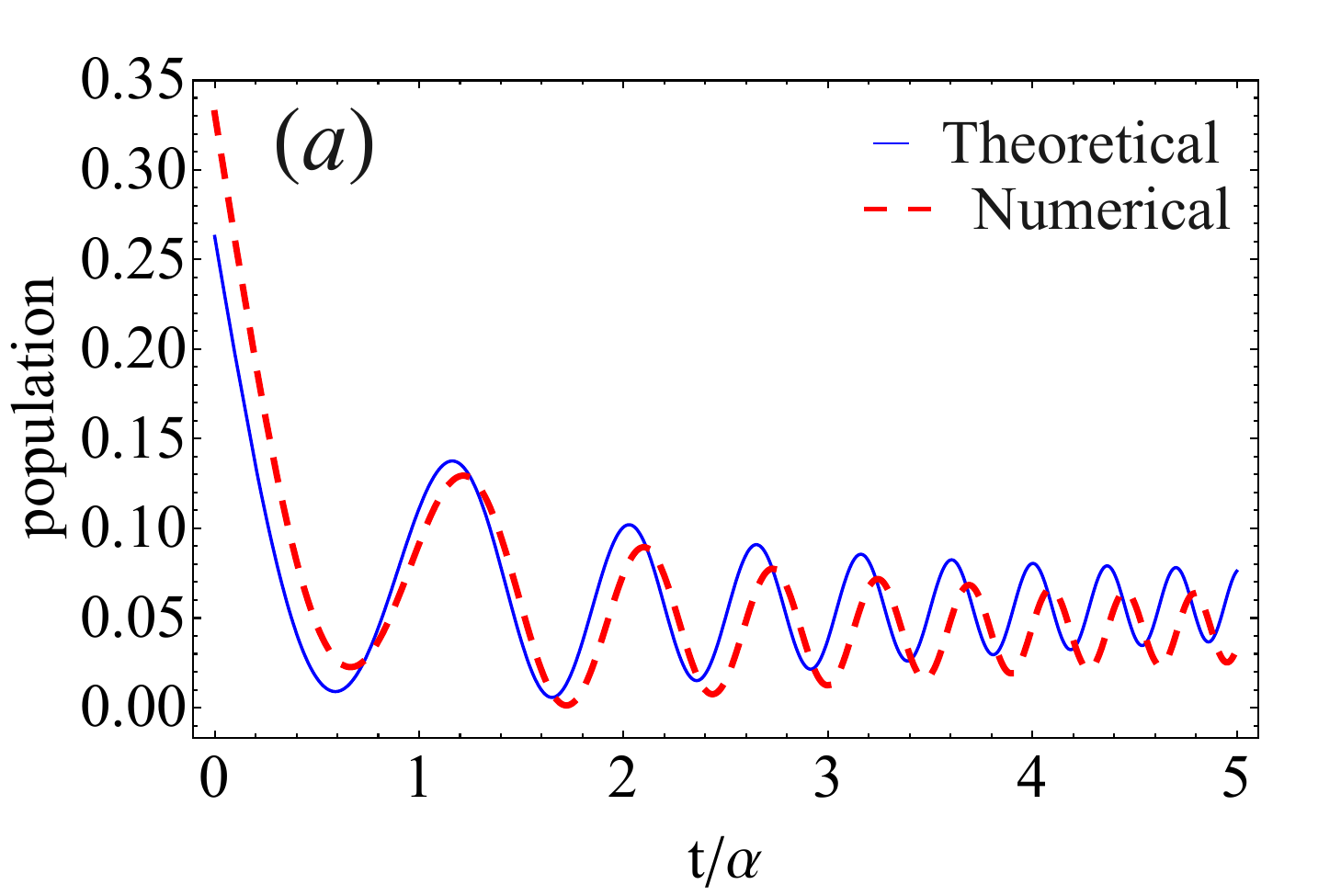}
	\includegraphics[width=0.23\textwidth, height =0.20\textwidth]{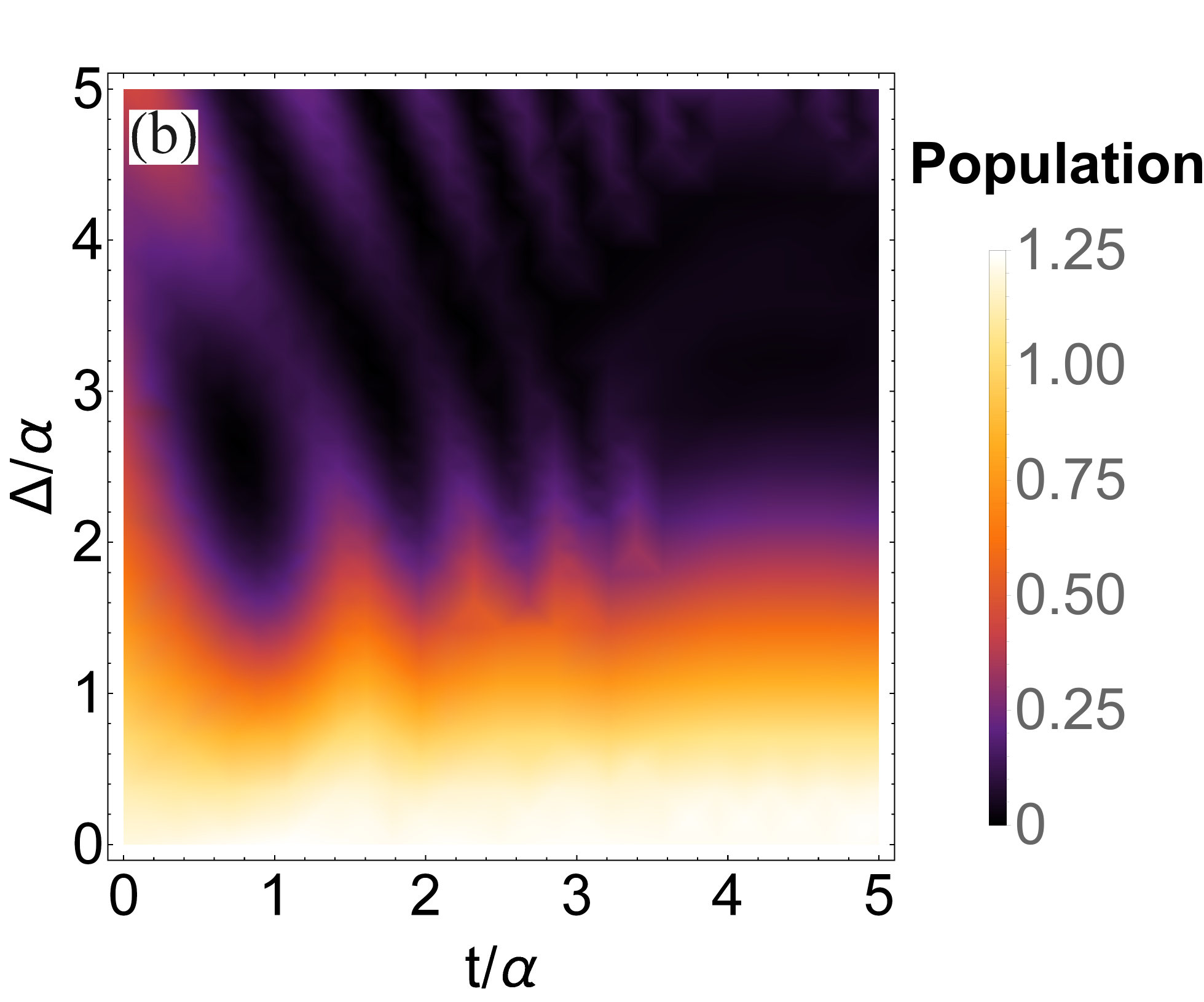}
	\caption{\small (Color online) (a) Similarity between theoretical and numerical simulations of the Landau-Zener model with imaginary coupling. (b) Interferogram of the Landau-Zener model with imaginary coupling. In this figure we have considered $\kappa = 0.3/, P = 4, \Delta > 0 $. Figure (b) shows how we can materialize the small dark and bright fringes using the jump $\kappa$. This alternate would be at the origin of the Landau-Zener interferometer. Our theoretical results have the same behavior as the numerical simulations.}\label{fig8}
\end{figure}

where the control parameter of the Landau-Zener model is $\lambda  = {\left( {i\Delta  + \kappa } \right)^2}/\alpha $.
At this level, the probailities amplitudes \cite{KenmoeMB} are given by:
\begin{equation}
	{C_{1}}\left( {x,{x_0}} \right) =  \frac{{{e^{i\pi /4}}}}{{\sqrt \lambda  }}\left( {{a_ + }{D_{ - i\lambda }}\left( {iz} \right) - a - {D_{ - i\lambda }}\left( { - iz} \right)} \right),
\end{equation} 

\begin{equation}
	{C_2}\left( {x,{x_0}} \right) = {a_ + }{D_{ - i\lambda  - 1}}\left( {iz} \right) - a - {D_{ - i\lambda  - 1}}\left( { - iz} \right).\label{5.4}
\end{equation}

The constants $a_ + $ and $a_ - $ are determined through \cite{Kammogne} using the initial conditions ${C_1}\left( {{x_0},{x_0}} \right) = 1$ and ${C_2}\left( {{x_0},{x_0}} \right) = 0$. The survival and transition probabilities are determined by:
\begin{equation}
	{P_{Sur}}\left( {x,{x_0}} \right) = {\mathop{\rm Re}\nolimits} {\left( {{C_1}\left( {x,{x_0}} \right)} \right)^2} + {\mathop{\rm Im}\nolimits} {\left( {{C_1}\left( {x,{x_0}} \right)} \right)^2},
\end{equation}

\begin{equation}
	{P_{Tra}}\left( {x,{x_0}} \right) = {\mathop{\rm Re}\nolimits} {\left( {{C_2}\left( {x,{x_0}} \right)} \right)^2} + {\mathop{\rm Im}\nolimits} {\left( {{C_2}\left( {x,{x_0}} \right)} \right)^2}.
\end{equation}

In Fig.\ref{fig8}, we show the analogy between our analytical and numerical results in (a) and present the interferogram in (b), which shows the presence of light beams with dark and bright transparent lines. Having presented all this, we now turn to the general conclusion of our work.

\section*{Conclusion}

The aim of this work was to study the effect of spontaneous emission on a tanh model. This effect was modeled by an imaginary coupling and a shift in the off-diagonal part of the Hamiltonian. Its dynamics are studied using the Schr\"odinger equation, which enabled us to determine its transition probability and energy.

As a result, we observe in Fig.\ref{fig2} that the population is characterized by a light beam represented by transparent dark and light rays when the coupling $\Delta$  is greater than zero. However, when the coupling $\Delta = 0$,  we notice the presence of more visible dark and light rays. In Fig.\ref{fig3}, the population is represented by a light beam, which is more condensed, with a large dark part and a small bright part for high values of the shift. Fig.\ref{fig4} shows a population that exceeds 1 due to a large amplitude value, and its interferogram shows a more condensed light beam.

Due to the effect of spontaneous emission on our model, an imaginary coupling and a shift are formed, which give our system a non-Hermitian character and are represented by the real and imaginary parts of the energy. We have established the necessary condition for observing the transmission of information in the energy diagram of the real part. This condition shows that when $P > 4$, the two energy levels communicate. However, when this condition is violated, quantum information is no longer transmitted. Regarding the energy diagram of the imaginary part, we observe that time and sweep velocity dictate the order and disorder in the system, and this influences the dynamics of our model.

In Figs.\ref{fig7}-\ref{fig8},  we show that our model has similarities with the Rabi model in the short-time approximation and the Landau-Zener model in the fast-time approximation. In this work, our theoretical results agree with numerical simulations. In future work, it would be interesting to study the effects of noise on this model, along with the effects of spontaneous emission.

\section*{Acknowledgments}

A. D. Kammogne thanked M. B. Kenmoe for his interesting scientific comments and the University of Dschang for its warm hospitality during this project.

\section*{Declaration}
This paper has received no funding or financial support from anyone.

\end{document}